\documentclass[usenatbib]{mnras}
\pdfoutput=1
\usepackage{graphicx}
\usepackage{ltablex}
\usepackage{supertabular}

\newcommand{\beq}{\begin{equation}}
\newcommand{\eeq}{\end{equation}}
\newcommand{\beqary}{\begin{eqnarray}}
\newcommand{\eeqary}{\end{eqnarray}}
\newcommand{\Lp}{L_{\rm p}}
\newcommand{\Lb}{L_{\rm b}}
\newcommand{\gmp}{\Gamma_{\rm p}}
\newcommand{\bs}{B_{\rm s}}
\newcommand{\blc}{B_{\rm lc}}
\newcommand{\edot}{\dot{E}}
\newcommand{\pdot}{\dot{P}}
\newcommand{\dV}{\Delta V}
\newcommand{\dVh}{\Delta V/h}
\newcommand{\epc}{E_{\rm pc}}

\title[Non-thermal X-ray emission from PSRs]{Observational connection of non-thermal X-ray emission from pulsars 
with their  timing properties and thermal emission}
\author[Chang et al.]{Hsiang-Kuang Chang,$^{1,2,3}$ \thanks{E-mail: hkchang@mx.nthu.edu.tw}
Jr-Yue Hsiang,$^{1}$ 
Che-Yen Chu,$^{1}$
Yun-Hsin Chung,$^{1}$
Tze-Hsiang Su,$^{2}$
\newauthor
Tzu-Hsuan Lin$^{2}$ and
Chien-You Huang$^{2}$
\\
$^{1}$Institute of Astronomy, National Tsing Hua University, Hsinchu 300044, Taiwan
\\
$^{2}$Department of Physics, National Tsing Hua University, Hsinchu 300044, Taiwan
\\
$^{3}$Center for Informatics and Computation in Astronomy (CICA), National Tsing Hua University, Hsinchu 300044, Taiwan
}

\begin{document}

\label{firstpage}
\date{Accepted XXX. Received YYY; in original form ZZZ}
\pubyear{2023}
\pagerange{\pageref{firstpage}--\pageref{lastpage}}
\maketitle

\begin{abstract}
The origin and radiation mechanisms of high energy emissions from pulsars have remained mysterious since their discovery.
Here we report, based on a sample of 68 pulsars, observational connection of non-thermal X-ray emissions from pulsars with their timing properties and thermal emissions, 
which may provide some constraints on theoretical modeling.
Besides strong correlations with the spin-down power $\edot$ and the magnetic field strength at the light cylinder $\blc$,
the non-thermal X-ray luminosity in 0.5 -- 8 keV,  $\Lp$, represented by the power-law component in the spectral model,
is  found to be strongly correlated with the highest possible electric field strength in the polar gap, $\epc$, of the pulsar.
The spectral power index $\gmp$ of that power-law component is also found, for the first time in the literature, 
to strongly correlate with $\edot$, $\blc$ and $\epc$, thanks to the large sample.
In addition, we found that $\Lp$ can be well described by
$\Lp\propto T^{5.96\pm 0.64}R^{2.24\pm 0.18}$, where $T$ and $R$ are the surface temperature and the emitting-region radius of the surface thermal emission,
represented by the black-body component in the spectral model.
 $\gmp$, on the other hand, 
can be well described only when timing variables are included, 
and the relation is
$\gmp = \log(T^{-5.8\pm 1.93}R^{-2.29\pm 0.85}P^{-1.19\pm 0.88}\pdot^{0.94\pm 0.44})$ plus a constant.
These relations strongly suggest the existence of connections between surface thermal emission and electron-positron pair production in pulsar magnetospheres.
\end{abstract}


\section{Introduction}

Pulsars are fast rotating, strongly magnetized neutron stars.
They are detected through their electromagnetic emissions, from radio waves to very high energy gamma rays.
These emissions carry rich information about neutron stars, their magnetosphere and likely also their wind zone.
After more than half a century since pulsar's discovery, however, except for the thermal component seen in X-rays,
how and from where these emissions come remain in debate.
Besides the unsolved issue of pulsar's coherent radio emission mechanism,
theoretical modeling to understand high-energy non-coherent emissions has been evolving.
More recent models for high-energy emissions from pulsars include, for example,
the striped-wind model \citep{petri13,petri15}, 
the extended-slot-gap and equatorial-current-sheet model \citep{harding18,harding21,barnard22},
the synchro-curvature model \citep{torres18,torres19,iniguez22}, 
and the non-stationary outer-gap model \citep{takata16,takata17}.

In all these models, non-thermal X-ray emissions are considered to come from synchrotron radiation of 
secondary electron-positron pair plasmas. These pairs are created in different location in different models.
In the outer-gap model, pairs are created in the outer gap close to the null surface through two-photon ($\gamma+\gamma$) pair production,
mainly from the collision of gamma-ray photons emitted by inwardly moving charged particles and X-ray photons from surface thermal emission \citep{cheng86,takata16}.
The resultant non-thermal X-ray emission from the synchrotron radiation of these pairs apparently depends on the local magnetic fields and the energy distribution of these pairs.
Therefore, one can expect that surface thermal emission plays an essential role in determining the properties of non-thermal X-rays.
In the slot-gap model, pairs are created close to the stellar surface above the magnetic polar cap through one-photon ($\gamma+B$) pair production cascades,
due to the interaction between  gamma-ray photons emitted by charged particles accelerated in the gap above the polar cap and the strong magnetic fields there \citep{harding18}.
The energetics of charged particles accelerated in the polar gap may be affected by surface thermal emission through the inverse Compton scattering between them 
\citep{chang95,sturner95,harding98,harding02,timokhin19}.
In such a case, one can also expect that surface thermal emission plays some role in determining the energetics and amount of the created pairs.
That may have some imprints in 
their synchrotron radiation in the slot gap or current sheets in the wind zone, that is, the observed non-thermal X-rays.
The connection between thermal and non-thermal X-rays is actually expected in different models, although not yet further elaborated in any of them.
The exact contents of the connection, when reliably established, can be a new diagnostic tool to constrain theoretical models.

In the past, almost all the efforts were focused on looking for the correlation between pulsars' X-ray luminosity and timing properties.
It was generally found that pulsars' X-ray luminosity, $L_{\rm X}$, is strongly correlated with  their spin-down power, $\edot$.
Their relationship, when expressed in a power law, is $L_{\rm X}\propto\edot^\alpha$ with $\alpha$ ranging from  0.92 to 1.87 (see \citet{hsiang21} for a brief review).   
In those earlier studies, however, $L_{\rm X}$ may be a mixture of thermal and non-thermal components 
and in some of them may also include contribution from associated pulsar wind nebulae. 
In all the cases the reduced $\chi^2$ of the power-law fitting for $L_{\rm X}$ and $\edot$ is always about 3 or larger.
Apparently the relationship is not tight.
Recently it was pointed out that $L_{\rm X}$ is also strongly correlated with $\blc$, 
the magnetic field strength at the light cylinder, at a significance level similar to that of the case with $\edot$
\citep{malov19,hsiang21}.
The scatter in the best fit of $L_{\rm X}$ as a power-law function of $\blc$ is also large, again similar to the case with $\edot$.
The comparable importance of $\edot$ and $\blc$ in correlating $L_{\rm X}$ suggests a scenario that $\edot$ is not the only major factor.
All the timing variables, such as $\edot$, $\blc$, characteristic age $\tau$ and the magnetic field at stellar surface $\bs$, are just a function of $P$ and $\pdot$,
the pulsar spin period and its time derivative.
A two-variable fitting of $L_{\rm X}$ seems desired.
However, efforts to fit $L_{\rm X}$ directly with $P$ and $\pdot$ did not yield any improvements.
The reduced $\chi^2$ is similarly large  \citep{possenti02,hsiang21}.

The large scatter, i.e., the large reduced $\chi^2$, in the aforementioned relationship fitting 
may be inevitable.
The geometry factors like the inclination angle between directions of the pulsar spin and magnetic moment and
the viewing angle between that of the spin and observer are not taken into account when
inferring luminosities from measured fluxes. It therefore goes  with the assumption of an isotropic emission, which is not expected to be true in general.
On the other hand, it may also indicate that there could be other non-negligible factors missing in the study.
As discussed above, surface thermal emission is expected to play some role in the radiation of non-thermal X-rays.
It is therefore tempting to look for possible relationships between spectral properties of non-thermal and thermal X-ray emissions, besides timing variables.
That is to say, there could be a better-defined fundamental plane of  non-thermal X-ray properties, 
such as its luminosity and the photon index of the corresponding power-law component, 
in a space spanned not only by $P$ and $\pdot$ but also by  thermal X-ray properties, such as the temperature and the emitting region radius.

In this paper we report the results of such an effort with a sample of 68 pulsars, 
among which 32 have both thermal and non-thermal X-ray emissions.
Sample collection is described in Section 2.
For the whole sample of 68 pulsars, we first investigate the correlation between non-thermal X-ray spectral properties and timing variables,
similar to what have been reported in the literature, but with some new findings. This section also serves as a consistency check between earlier results 
in the literature and ours.
These are described in Section 3.  
We then present the search for the fundamental plane in a four-dimensional space for the 32 pulsars with thermal emissions in Section 4.
Discussions and conclusions are in Section 5.
     
\section{Sample collection}
 \begin{table*}
\renewcommand{\arraystretch}{1.7}
\caption{X-ray spectral properties of 68 pulsars. $\Lp$ and $\gmp$ are the power-law-component luminosity (0.5 -- 8 keV) and photon index. 
The surface temperature $T$ (expressed together with the Boltzmann constant $k$ in units of eV) and  the emitting-region radius $R$ are 
the measured values obtained from spectral fitting. They are the so-called temperature and radius measured at infinity (far away from the neutron star).
References: (1) This work; (2) \citet{J0108-1431}; (3) \citet{J0357+3205};  (4) \citet{J0358+5413}; (5) \citet{li08}; (6) \citet{J0537-6910};
(7) \citet{J0540-6919}; (8) \citet{J0630-2834}; (9) \citet{J0633+0632}; (10) \citet{J0633+1746}; (11) \citet{J0659+1414}; (12) \citet{J0826+2637}; (13) \citet{J0835-4510}; (14) \citet{J0922+0638}; (15) \citet{J0946+0951}; (16) \citet{J0953+0755}; (17) \citet{J1016-5857}; (18) \citet{J1048-5832}; (19) \citet{J1057-5226}; (20) \citet{J1101-6101}; (21) \citet{J1136+1551}; (22) \citet{J1154-6250}; (23) \citet{J1357-6429}; (24) \citet{J1418-6058};  (25) \citet{J1456-6843}; (26) \citet{J1459-6053}; (27) \citet{J1509-5850}; (28) \citet{J1617-5055}; (29) \citet{J1709-4429}; (30) \citet{J1740+1000};  (31) \citet{J1741-2054}; (32) \citet{J1747-2958}; (33) \citet{J1801-2451}; (34) \citet{J1809-1917}; (35) \citet{J1811-1925}; (36) \citet{J1813-1749_2}; (37) \citet{J1826-1256}; (38) \citet{J1826-1334}; (39) \citet{J1836+5925};
(40) \citet{coti20}; (41) \citet{J1932+1059}; (42) \citet{J1952+3252}; (43) \citet{J1957+5033}; (44) \citet{J2021+3651}; (45) \citet{J2021+4026};  (46) \citet{J2055+2539}; (47) \citet{J2225+6535}; (48) \citet{J2337+6151}.
$^\dagger$PSR J0946+0951 is a radio and X-ray mode-switching pulsar. It is treated as 2 distinct sources in our study.} 
\label{tab:sp}
\begin{tabular}{lllllll}
\hline
\hline
Pulsar name & $\Lp$ (erg/s)  & $\gmp$ & $kT$ (eV)   &$R$ (km) & References \\
\hline

J0007+7303 
    & $8.82^{+10.0}_{-5.97} \times 10^{30}$ & $1.71^{+0.31}_{-0.28}$ 
    & $-$   & $-$  
    & 1\\
J0108$-$1431 
    & $4.40^{+8.72}_{-3.36} \times 10^{28}$ & $3.1^{+0.5}_{-0.2}$ 
    & $-$   & $-$
    & 2\\
J0205+6449 
    & $1.06^{+1.07}_{-0.69} \times 10^{33}$ & $1.47^{+0.03}_{-0.03}$ 
    & $167^{+31}_{-19}$     & $1.30^{+1.95}_{-0.92}$
    & 1\\
J0357+3205 
    & $3.44^{+3.55}_{-2.42} \times 10^{30}$ & $2.28^{+0.17}_{-0.16}$ 
    & $94^{+12}_{-9}$   & $0.76^{+1.03}_{-0.49}$ 
    & 3\\
J0358+5413 
    & $4.37^{+7.45}_{-3.42} \times 10^{30}$ & $1.45^{+0.21}_{-0.24}$ 
    & $160^{+50}_{-40}$     & $2.43^{+7.33}_{-1.53}$  
    & 4\\
J0534+2200 
    & $1.32^{+1.89}_{-0.94} \times 10^{36}$ & $1.63^{+0.09}_{-0.09}$ 
    & $-$   & $-$
    & 5\\
J0537$-$6910 
    & $5.02^{+5.10}_{-3.26} \times 10^{35}$ & $1.73^{+0.11}_{-0.06}$ 
    & $-$   & $-$
    & 6\\
J0540$-$6919 
    & $3.01^{+3.18}_{-1.98} \times 10^{36}$ & $0.78^{+0.09}_{-0.09}$ 
    & $-$   & $-$  
    & 5,7\\
J0630$-$2834 
    & $2.22^{+3.67}_{-1.58} \times 10^{29}$ & $2.27^{+0.23}_{-0.13}$ 
    & $250^{+50}_{-40}$     & $0.02^{+0.02}_{-0.01}$ 
    & 8\\
J0633+0632 
    & $1.31^{+3.76}_{-1.19} \times 10^{31}$ & $1.6^{+0.6}_{-0.6}$ 
    & $105^{+23}_{-18}$     & $3.26^{+11.59}_{-2.04}$ 
    & 9\\ 
J0633+1746 
    & $1.48^{+1.46}_{-1.02} \times 10^{30}$ & $1.47^{+0.06}_{-0.07}$
    & $72^{+4}_{-4}$    & $2.13^{+1.57}_{-1.42}$   
    & 10\\
J0659+1414 
    & $1.49^{+2.79}_{-1.06} \times 10^{30}$  & $2.30^{+0.68}_{-0.57}$ 
    & $66^{+3}_{-6}$    & $0.41^{+0.32}_{-0.19}$ 
    & 11\\
J0826+2637 
    & $3.35^{+3.83}_{-2.26} \times 10^{29}$  & $2.50^{+0.17}_{-0.17}$ 
    & $290^{+30}_{-30}$     & $0.02^{+0.01}_{-0.01}$  
    & 12\\
J0835-4510 
    & $5.27^{+6.54}_{-3.64} \times 10^{32}$ & $2.7^{+0.4}_{-0.4}$ 
    & $128^{+3}_{-3}$   & $2.35^{+1.25}_{-1.08}$ 
    & 13\\
J0922+0638 
    & $1.53^{+2.90}_{-1.26} \times 10^{30}$ & $2.2^{+0.6}_{-0.6}$ 
    & $110^{+20}_{-30}$     & $0.21^{+0.33}_{-0.13}$  
    & 14\\
J0946+0951$_{\_B}$ 
    & $3.71^{+5.90}_{-2.84} \times 10^{29}$ & $2.2^{+0.2}_{-0.3}$ 
    & $210^{+20}_{-20}$     & $0.04^{+0.03}_{-0.02}$ 
    & 15\\
J0946+0951$_{\_Q}$ 
    & $9.02^{+11.4}_{-6.34} \times 10^{29}$ & $2.6^{+0.2}_{-0.1}$ 
    & $300^{+20}_{-20}$     & $0.03^{+0.02}_{-0.01}$ 
    & 15\\
J0953+0755 
    & $6.48^{+7.74}_{-4.39} \times 10^{29}$ & $1.92^{+0.14}_{-0.12}$ 
    & $-$   & $-$  
    & 16\\
J1016$-$5857 
    & $7.20^{+7.25}_{-4.67} \times 10^{32}$ & $1.08^{+0.08}_{-0.08}$ 
    & $-$   & $-$  
    & 17\\
J1023$-$5746 
    & $2.27^{+2.21}_{-1.46} \times 10^{33}$ & $0.84^{+0.42}_{-0.42}$ 
    & $-$   & $-$   
    & 1\\
J1048$-$5832 
    & $2.43^{+3.08}_{-1.69} \times 10^{31}$ & $1.5^{+0.3}_{-0.3}$ 
    & $-$   & $-$  
    & 18\\
J1057$-$5226 
    & $8.40^{+16.3}_{-0.69} \times 10^{28}$ & $1.9^{+0.2}_{-0.2}$   
    & $85^{+6}_{-6}$    & $0.73^{+0.48}_{-0.36}$ 
    & 19\\
J1101$-$6101 
    & $3.53^{+3.53}_{-2.29} \times 10^{33}$ & $1.10^{+0.06}_{-0.06}$ 
    & $-$   & $-$ 
    & 20\\
J1112$-$6103 
    & $9.97^{+14.5}_{-6.99} \times 10^{31}$ & $1.09^{+0.65}_{-0.51}$ 
    & $-$   & $-$ 
    & 1\\
J1124$-$5916
    & $1.23^{+2.03}_{-0.89} \times 10^{33}$ & $0.94^{+0.07}_{-0.08}$ 
    & $460^{+70}_{-80}$     & $0.35^{+0.31}_{-0.18}$  
    & 1\\
J1136+1551 
    & $8.06^{+12.2}_{-6.40} \times 10^{28}$ & $2.0^{+1.3}_{-1.2}$ 
    & $190^{+60}_{-52}$     & $0.02^{+0.03}_{-0.01}$ 
    & 21\\
J1154$-$6250 
    & $1.58^{+2.31}_{-1.16} \times 10^{30}$ & $3.1^{+0.5}_{-0.4}$ 
    & $-$   & $-$ 
    & 22\\
J1357$-$6429 
    & $6.20^{+12.8}_{-4.63} \times 10^{31}$ & $1.72^{+0.55}_{-0.63}$ 
    & $140^{+60}_{-40}$     & $2.52^{+18.64}_{-2.10}$ 
    & 23\\
J1418$-$6058 
    & $5.51^{+7.45}_{-3.92} \times 10^{31}$ & $1.5^{+0.4}_{-0.4}$ 
    & $-$   & $-$ 
    &24\\
J1420$-$6048 
    & $3.74^{+4.19}_{-2.51} \times 10^{32}$ & $0.63^{+0.33}_{-0.34}$ 
    & $-$   & $-$ 
    & 1\\
\hline
\end{tabular}
\end{table*}  
 \begin{table*}
 \renewcommand{\arraystretch}{1.7}
\contcaption{}
\begin{tabular}{lllllll}
\hline
\hline
Pulsar name & $\Lp$ (erg/s)  & $\gmp$ & $kT$ (eV)   &$R$ (km) & References \\
\hline   
J1456$-$6843 
    & $1.02^{+1.68}_{-0.74} \times 10^{30}$ & $2.4^{+0.4}_{-0.3}$ 
    & $-$   & $-$  
    & 25\\
J1459$-$6053 
    & $2.87^{+3.23}_{-1.93} \times 10^{31}$ & $1.08^{+0.18}_{-0.17}$ 
    & $-$   & $-$
    & 26\\
J1509$-$5850 
    & $7.14^{+7.68}_{-4.72} \times 10^{31}$ & $1.90^{+0.12}_{-0.12}$ 
    & $-$   & $-$  
    & 27\\
J1513$-$5908 
    & $6.42^{+6.33}_{-4.14} \times 10^{34}$ & $1.35^{+0.03}_{-0.03}$ 
    & $119^{+9}_{-10}$  & $34.82^{+29.28}_{-21.19}$ 
    & 1\\
J1617$-$5055 
    & $9.62^{+9.77}_{-6.26} \times 10^{33}$ & $1.14^{+0.06}_{-0.06}$    
    & $-$   & $-$  
    & 28\\
J1709$-$4429 
    & $1.82^{+1.91}_{-1.19} \times 10^{32}$ & $1.62^{+0.20}_{-0.20}$ 
    & $172^{+15}_{-14}$     & $2.08^{+2.77}_{-1.35}$ 
    & 29\\
J1718$-$3825 
    & $4.76^{+5.83}_{-3.26} \times 10^{31}$ & $1.62^{+0.33}_{-0.30}$ 
    & $-$   & $-$ 
    & 1\\
J1732$-$3131 
    & $3.06^{+3.79}_{-2.11} \times 10^{30}$ & $1.14^{+0.35}_{-0.22}$ 
    & $-$   & $-$  
    & 1\\   
J1740+1000 
    & $4.96^{+6.90}_{-3.57} \times 10^{30}$ & $1.80^{+0.17}_{-0.17}$ 
    & $83^{+5}_{-5}$    & $4.86^{+3.58}_{-2.43}$  
    & 30\\
J1741$-$2054 
    & $4.16^{+4.42}_{-2.74} \times 10^{30}$ & $2.66^{+0.06}_{-0.06}$ 
    & $60^{+2}_{-2}$    & $5.10^{+3.51}_{-2.58}$ 
    & 31\\
J1747$-$2958 
    & $1.48^{+1.47}_{-0.96} \times 10^{33}$ & $1.55^{+0.04}_{-0.04}$ 
    & $-$   & $-$  
    & 32\\
J1801$-$2451 
    & $1.54^{+1.74}_{-1.03} \times 10^{33}$ & $1.6^{+0.6}_{-0.5}$ 
    & $-$   & $-$  
    & 33\\
J1803$-$2137 
    & $5.93^{+8.59}_{-4.14} \times 10^{31}$ & $2.0^{+0.4}_{-0.3}$ 
    & $-$   & $-$  
    & 18\\
J1809$-$1917 
    & $7.26^{+9.38}_{-5.09} \times 10^{31}$ & $1.28^{+0.15}_{-0.15}$ 
    & $190^{+30}_{-30}$     & $0.53^{+0.81}_{-0.33}$ 
    & 34\\
J1809$-$2332 
    & $1.42^{+1.60}_{-0.95} \times 10^{31}$ & $1.36^{+0.16}_{-0.15}$ 
    & $178^{+23}_{-20}$     & $0.43^{+0.59}_{-0.27}$  
    &  1\\
J1811$-$1925 
    & $6.97^{+7.27}_{-4.57} \times 10^{33}$ & $0.97^{+0.32}_{-0.39}$ 
    & $-$   & $-$  
    & 35\\
J1813$-$1246 
    & $6.78^{+6.63}_{-4.36} \times 10^{32}$ & $0.85^{+0.03}_{-0.03}$ 
    & $-$ & $-$ 
    & 1\\
J1813$-$1749 
    & $8.21^{+9.03}_{-7.16} \times 10^{33}$ & $1.44^{+0.44}_{-0.42}$ 
    & $-$   & $-$  
    &  36\\
J1826$-$1256 
    & $2.32^{+2.84}_{-1.59} \times 10^{31}$ & $0.92^{+0.25}_{-0.24}$ 
    & $-$   & $-$ 
    & 37\\
J1826$-$1334 
    & $4.77^{+6.91}_{-3.47} \times 10^{31}$ & $2.43^{+0.53}_{-0.40}$ 
    & $-$   & $-$ 
    & 38\\
J1833$-$1034 
    & $3.75^{+5.25}_{-2.59} \times 10^{33}$ & $1.50^{+0.33}_{-0.31}$ 
    & $-$   & $-$ 
    & 1\\
J1836+5925 
    & 3.18$^{+4.75}_{-2.35} \times 10^{29}$ & $2.1^{+0.3}_{-0.3}$ 
    & $45^{+10}_{-8}$   & $2.36^{+5.04}_{-1.71}$ 
    &  39\\
J1838$-$0537 
    & $1.12^{+4.54}_{-0.83} \times 10^{31}$ & $1.2^{+1.0}_{-1.0}$ 
    & $-$   & $-$  
    & 40\\
J1838$-$0655 
    & $3.04^{+3.38}_{-2.01} \times 10^{34}$ & $0.82^{+0.17}_{-0.17}$ 
    & $-$   & $-$  
    &  1\\
J1856+0113 
    & $1.50^{+2.12}_{-1.04} \times 10^{32}$ & $2.25^{+0.28}_{-0.26}$ 
    & $-$   & $-$ 
    &  1\\
J1907+0602 
    & $4.73^{+6.20}_{-1.68} \times 10^{31}$ & $0.76^{+1.07}_{-0.34}$ 
    & $-$   & $-$  
    & 1\\
J1930+1852
    & $8.62^{+9.58}_{-5.81} \times 10^{33}$ & $0.76^{+0.20}_{-0.25}$ 
    & $365^{+143}_{-96}$    & $0.84^{+2.39}_{-0.61}$  
    & 1\\
J1932+1059
    & $1.03^{+3.30}_{-0.88} \times 10^{30}$ & $1.73^{+0.46}_{-0.66}$ 
    & $300^{+20}_{-30}$     & $0.03^{+0.01}_{-0.01}$  
    & 41\\
J1952+3252
    & $2.91^{+3.28}_{-1.95} \times 10^{33}$ & $1.63^{+0.05}_{-0.05}$ 
    & $130^{+20}_{-20}$     & $3.30^{+4.26}_{-2.04}$  
    & 42\\
J1957+5033
    & $6.49^{+7.74}_{-4.43} \times 10^{30}$ & $1.76^{+0.11}_{-0.11}$ 
    & $54^{+5}_{-7}$    & $5.53^{+19.81}_{-5.23}$  
    & 43\\
J2021+3651
    & $1.86^{+2.15}_{-1.26} \times 10^{31}$ & $1.73^{+1.15}_{-1.02}$ 
    & $159^{+17}_{-21}$     & $1.25^{+1.52}_{-0.68}$  
    & 44\\
J2021+4026
    & $1.23^{+2.08}_{-0.88} \times 10^{31}$ & $1.0^{+0.7}_{-0.8}$ 
    & $210^{+30}_{-30}$     & $0.39^{+1.34}_{-0.39}$   
    & 45\\
J2022+3842
    & $7.67^{+9.63}_{-5.24} \times 10^{33}$ & $1.23^{+0.22}_{-0.22}$ 
    & $222^{+76}_{-37}$     & $5.23^{+16.50}_{-4.56}$  
    & 1\\
J2043+2740 
    & $2.85^{+5.31}_{-2.30} \times 10^{30}$ & $3.1^{+1.1}_{-0.6}$ 
    & $-$   & $-$ 
    & 16\\
J2055+2539 
    & $1.15^{+1.28}_{-0.77} \times 10^{30}$ & $2.36^{+0.14}_{-0.14}$ 
    & $-$   & $-$  
    & 46\\
J2225+6535
    & $1.71^{+2.04}_{-1.21} \times 10^{30}$ & $1.70^{+0.46}_{-0.23}$ 
    & $-$   & $-$ 
    & 47\\
J2229+6114
    & $8.29^{+9.01}_{-5.49} \times 10^{32}$ & $1.39^{+0.11}_{-0.12}$ 
    & $126^{+19}_{-13}$     & $5.17^{+10.13}_{-3.99}$  
    & 1\\
J2337+6151 
    & $8.44^{+8.89}_{-6.88} \times 10^{30}$ & $2.2^{+3.0}_{-1.4}$  
    & $109^{+30}_{-50}$     & $0.38^{+0.68}_{-0.21}$  
    & 48 \\

\hline
\end{tabular}
\end{table*}


We collected, from the literature, all the pulsars with detected X-ray emissions.
In this effort, among many different references, we benefited very much from the compiled tables in \citet{li08}, \citet{malov19}, \citet{coti20} and \citet{potekhin20}.
We excluded red-back and black-widow systems as  their X-ray emissions are from an intrabinary shock (e.g. \citet{kandel19}),
the two  pulsars with magnetar-like bursts, that is, PSR J1119-6127 \citep{blumer17,dai18,archibald18} and
PSR J1846-0258 \citep{kuiper18,reynolds18,temim19},
and those that are either too dim or with too short an exposure so that the number of source photons available
is too small to yield a meaningful spectral analysis.
Millisecond pulsars are not included in this study either.
Their non-thermal X-ray emissions 
may come from a way different from normal pulsars, because of their weak magnetic field (e.g. \citet{zhang03}).
We also found 10 pulsars with purely thermal X-ray emissions.
Their X-ray spectra can be well described by one or two blackbody components 
with signatures of lines in some of them.
Since there is no non-thermal X-ray emission detected from them, 
these 10 are excluded from the current study.
PSR J0943+0951 switches between two emission modes.
In its radio quiet mode (Q-mode) its X-ray emission is brighter and in the radio bright one (B-mode) its X-ray emission is fainter \citep{J0946+0951}.
The X-ray spectral properties are different in these two modes.
We treat it as two distinct pulsars.
In total, there are 68 pulsars in the sample.

We scrutinized over these 68 pulsars to see whether a spectral model consisting of
a power-law (PL) component and a blackbody (BB) component can well describe the observed X-ray spectrum,
either from the literature or, when not available, from the spectral analysis conducted in this work.
32 among the 68 pulsars were found to be well described by such a PL+BB model,
based on the criterion that the corresponding $p$-value (that is, the corresponding random probability) of the reduced $\chi^2$ of the best-fit spectral model is larger than 0.05. 
A single PL model fits the other 36 pulsars well. 
Similar to the approach taken in \citet{hsiang21}, we mainly used Chandra data in the energy range from 0.5 keV to 8 keV.
This energy range is the most common one employed in the spectral analysis we found in the literature, 
in particular for those based on Chandra data.
When fluxes and uncertainties in the literature were reported in a different energy range, 
we compute the corresponding flux in 0.5 -- 8 keV and  estimate the uncertainty with the published value scaled by the flux ratio of the two energy ranges.
To covert fluxes into luminosities, distances to the sources are needed.
We adopt the best-estimate distances provided in the ATNF pulsar catalog\footnote{https://www.atnf.csiro.au/research/pulsar/psrcat/}
for the conversion. 
Following \citet{possenti02}, \citet{li08} and \citet{hsiang21}, we also adopt an uncertainty of $\pm$40\% in distances.
Four pulsars,  PSR J0633+1746, PSR J0659+1414, PSR J1057-5226 and PSR J1740+1000, actually require a
PL+2BB model to yield an acceptable fit.
To have a uniform comparison base,  a flux-weighted average of the temperature and the emitting-region radius of the two blackbody components
is employed to describe their thermal emission.
All the spectral properties of these 68 pulsars are listed in Table \ref{tab:sp}.
The distances taken from ATNF pulsar catalog and the pulsar timing properties are in Table \ref{tab:tm}.

\begin{figure*}
\includegraphics[width=16cm]{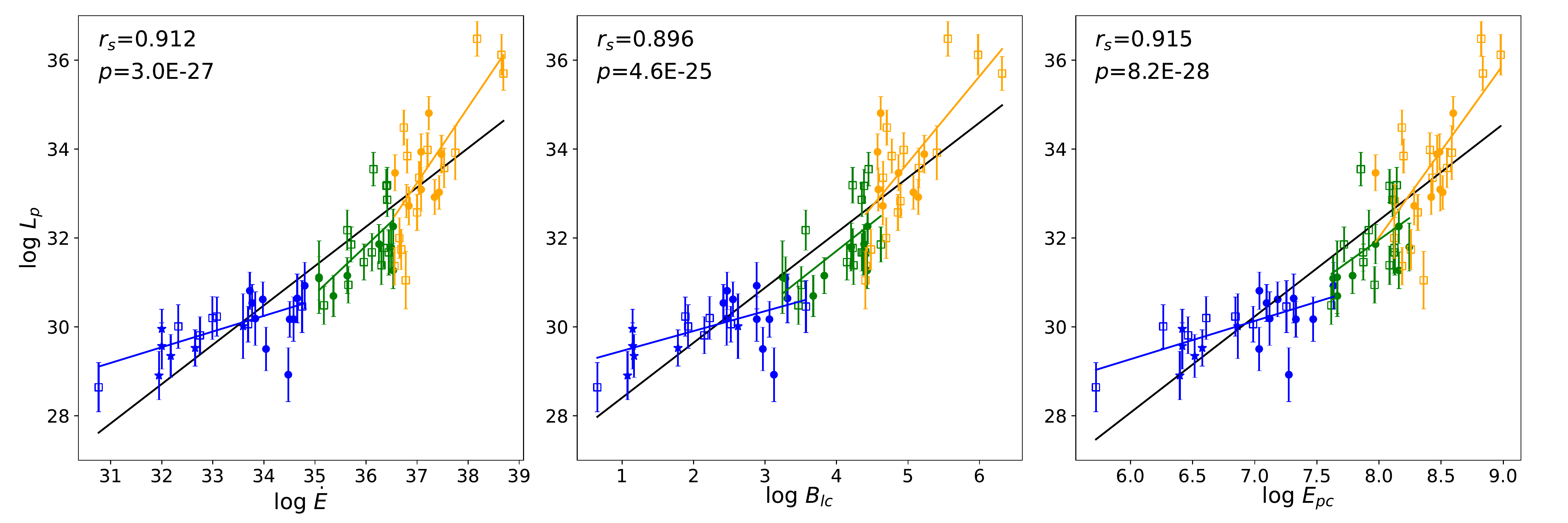}
\caption{The non-thermal X-ray luminosity $\Lp$ versus $\edot$, $\blc$ and $\epc$.
In the legend, $r_s$ is the Spearman rank-order correlation coefficient and $p$ is the corresponding random probability for the whole sample of
68 pulsars, whose best linear fit is shown with the line in black  in each panel.
These are the three with the strongest correlations between $\Lp$ and timing variables.
Pulsars with observed thermal emissions are plotted with solid symbols.
Those six with  a hot spot are plotted with a star symbol.
Three spin-down-power groups are distinguished in different colors.
The best linear fits for each group are plotted in corresponding colors. 
All the quantities are in gaussian units, as indicated in Table \ref{tab:tm}.
}
\label{fig:lptv}
\end{figure*}   
\begin{figure}
\includegraphics[width=8cm]{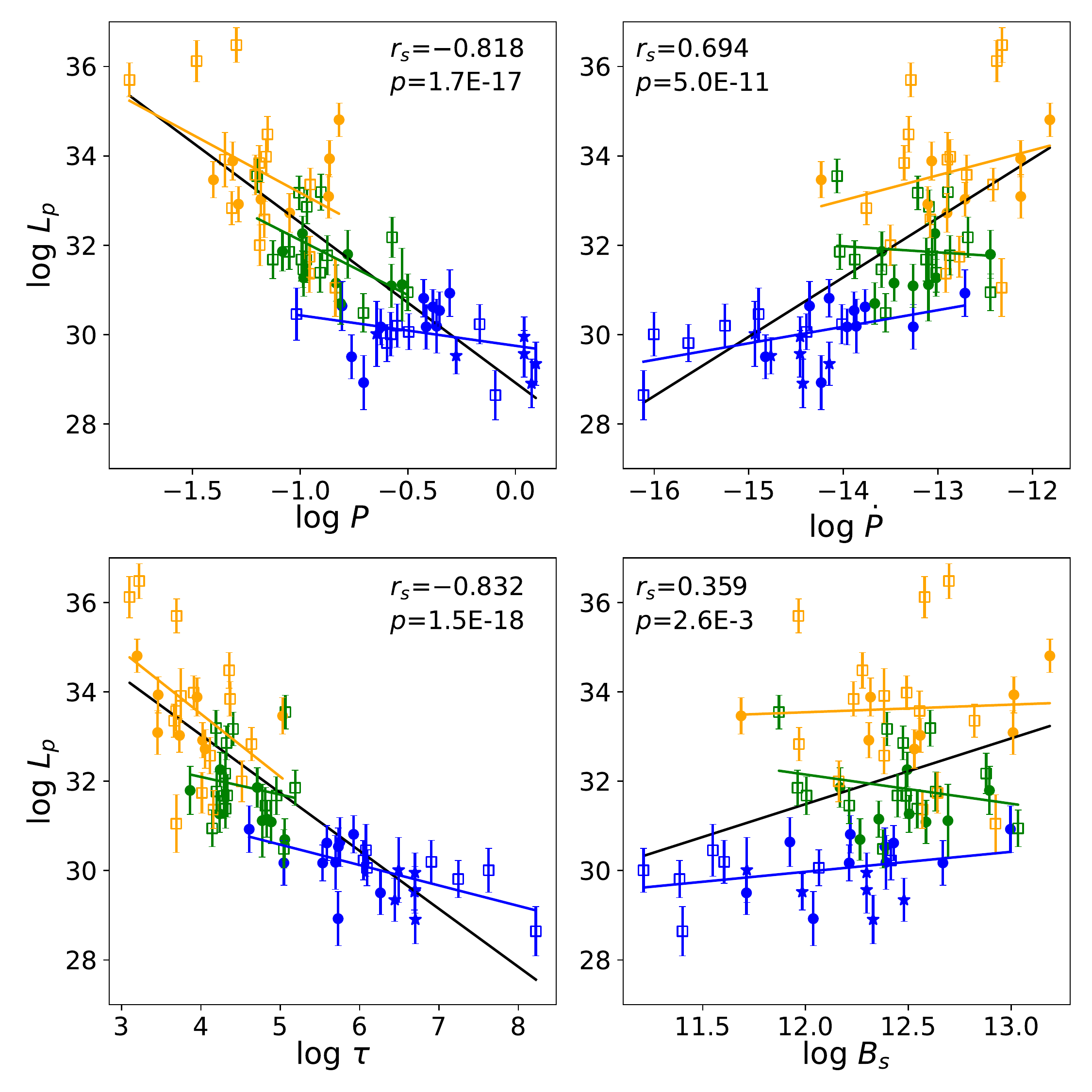}
\caption{The non-thermal X-ray luminosity $\Lp$ versus $P$, $\pdot$, $\tau$ and $B_{\rm s}$.
Legends are the same as in Figure \ref{fig:lptv}.
These are the four with weaker correlations between $\Lp$ and timing variables.
All the quantities are in gaussian units, except for the characteristic age $\tau$, as indicated in Table \ref{tab:tm}.
}
\label{fig:lptvx}
\end{figure}   

\begin{table*}
\renewcommand{\arraystretch}{1.38}
\caption{Distances and timing variables of the 67 pulsars employed in our analysis.}
\label{tab:tm}
\begin{tabular}{lllllllll}
\\
\hline
\hline
Source name & Distance  &  $P$  & $\dot{P}$ &  $\log \dot{E}$ &  $\log \tau$  
&  $\log B_{\rm s}$  & $\log B_{\rm lc}$  &$\log\epc$ \\
& (kpc) & (s) &(s/s) & (erg/s) & (yr) & (G)& (G)  & (statV/cm) \\
\hline
J0007+7303  &\ \ 1.40    &0.3159  & $3.60\times 10^{-13}$ & 35.65 & 4.14 & 13.03 & 3.51  & 7.96 \\
J0108$-$1431  &\ \ 0.21   &0.8076  & $7.70\times 10^{-17}$ &30.76  & 8.22& 11.40 & 0.65 & 5.72  \\
J0205+6449  &\ \ 3.20   & 0.0657 & $1.94\times 10^{-13}$ & 37.43 & 3.73 & 12.56 & 5.08 & 8.51 \\
J0357+3205  &\ \  0.84  &0.4441  & $1.30\times 10^{-14}$ & 33.77 & 5.73 & 12.39 & 2.42  & 7.10 \\
J0358+5413  &\ \ 1.00   & 0.1564 & $4.40\times 10^{-15}$ & 34.65 & 5.75 & 11.92 & 3.31 & 7.31 \\
J0534+2200  &\ \ 2.00   & 0.0330 & $4.21\times 10^{-13}$& 38.65 & 3.10 & 12.58 &5.98  & 8.98 \\
J0537$-$6910  &\ \ 49.70  & 0.0161 & $5.18\times 10^{-14}$&38.69  & 3.69 & 11.97 &6.32   & 8.84 \\
J0540$-$6919 &\ \ 49.70   & 0.0505 & $4.79\times 10^{-13}$& 38.18 & 3.22 & 12.70 & 5.56  & 8.82 \\
J0630$-$2834  &\ \ 0.32    &1.2444  & $7.12\times 10^{-15}$ & 32.18 & 6.44 & 12.48 & 1.17  & 6.52 \\
J0633+0632  &\ \ 1.36    &0.2974  & $7.96\times 10^{-14}$ & 35.08 & 4.77 & 12.69 & 3.24 & 7.66 \\
J0633+1746  &\ \ 0.19   & 0.2371 & $1.10\times 10^{-14}$  & 34.51 & 5.53 & 12.21 & 3.06  &7.33 \\
J0659+1414  &\ \ 0.29    & 0.3849 & $5.49\times 10^{-14}$& 34.58 & 5.05 & 12.67 & 2.88  & 7.47 \\
J0826+2637  &\ \ 0.50    &0.5307  & $1.71\times 10^{-15}$ & 32.65 & 6.69 & 11.98 & 1.78  &  6.58 \\
J0835$-$4510  &\ \ 0.28     & 0.0893 & $1.25\times 10^{-13}$& 36.84 & 4.05 & 12.53 &4.65  & 8.28 \\
J0922+0638  &\ \ 1.10    &0.4306  & $1.37\times 10^{-14}$ & 33.83 & 5.70 & 12.39 & 2.46  &  7.12\\
J0946+0951  &\ \ 0.89    &1.0977  & $3.49\times 10^{-15}$ & 32.00 & 6.70 & 12.30 & 1.15  & 6.42 \\
J0953+0755  &\ \ 0.26    &0.2531  & $2.30\times 10^{-16}$ & 32.75 & 7.24 & 11.39 & 2.15  &6.46  \\
J1016$-$5857  &\ \ 3.16    & 0.1074 & $8.08\times 10^{-14}$& 36.41 & 4.32 & 12.47 &4.35   & 8.11 \\
J1023$-$5746  &\ \ 2.08    &0.1115  & $3.84\times 10^{-13}$ & 37.04 & 3.66 & 12.82 & 4.65  &8.43  \\
J1048$-$5832  &\ \ 2.90    & 0.1237 & $9.61\times 10^{-14}$& 36.30 & 4.31 & 12.54 & 4.24  & 8.08 \\
J1057$-$5226  &\ \ 0.09    &0.1971  & $5.84\times 10^{-15}$ &  34.48& 5.73 & 12.04 & 3.12  & 7.27 \\
J1101$-$6101  &\ \ 7.00    & 0.0630 & $8.60\times 10^{-15}$& 36.15 & 5.06 & 11.87 & 4.45 &  7.85\\
J1112$-$6103  &\ \ 4.50    &0.0650  & $3.15\times 10^{-14}$ &36.65  & 4.51 & 12.16 & 4.69  & 8.12  \\
J1124$-$5916  &\ \ 5.00    & 0.1350 & $7.53\times 10^{-13}$& 37.08 & 3.45 & 13.01 & 4.59  & 8.49 \\
J1136+1551  &\ \ 0.37    &1.1879  & $3.73\times 10^{-15}$ & 31.94 & 6.70 & 12.33 & 1.08   &  6.40 \\
J1154$-$6350  &\ \ 1.36    & 0.2820  & $5.59\times 10^{-16}$ & 32.99 & 6.90 & 11.60 & 2.23  & 6.61 \\
J1357$-$6429  &\ \ 3.10     & 0.1661 & $3.60\times 10^{-13}$& 36.49 & 3.86 & 12.89 & 4.20  & 8.24 \\
J1418$-$6058  &\ \ 1.89    &0.1106  & $1.69\times 10^{-13}$ & 36.69 & 4.01 & 12.64 & 4.48  & 8.26 \\
J1420$-$6048  &\ \ 5.63    & 0.0682 & $8.32\times 10^{-14}$& 37.00 & 4.11 & 12.38 & 4.86  &  8.31 \\
J1456$-$6843  &\ \ 0.43    &0.2634  & $9.90\times 10^{-17}$ & 32.32 & 7.63 & 11.21 & 1.92 & 6.26 \\
J1459$-$6053  &\ \ 1.84   &0.1032  & $2.53\times 10^{-14}$ & 35.96 & 4.81 & 12.21 & 4.15  &  7.87 \\
J1509$-$5850  &\ \ 3.37    & 0.0889 & $9.17\times 10^{-15}$& 35.71 & 5.19 & 11.96 & 4.62  & 7.72 \\
J1513$-$5908  &\ \ 4.40   & 0.1516 & $1.53\times 10^{-12}$& 37.23 & 3.20 & 13.19 & 4.62  & 8.60 \\
J1617$-$5055  &\ \ 4.74    & 0.0694 & $1.35\times 10^{-13}$& 37.20 & 3.91 & 12.49 & 4.94  & 8.41 \\
J1709$-$4429  &\ \ 2.60    & 0.1020 & $9.30\times 10^{-14}$& 36.53 & 4.24 & 12.49 & 4.43  & 8.16 \\
J1718$-$3825  &\ \ 3.49    & 0.0747 & $1.32\times 10^{-14}$& 36.11 & 4.95 & 12.00 & 4.35  & 7.87 \\
J1732$-$3131  &\ \ 0.64    & 0.1965 & $2.80\times 10^{-14}$ & 35.18 & 5.05 & 12.38 & 3.47  &  7.62\\
J1740+1000  &\ \ 1.23  & 0.1541  & $2.15\times 10^{-14}$ & 35.36 & 5.06 & 12.26 & 3.67  & 7.66 \\
J1741$-$2054  &\ \ 0.30    &0.4137  & $1.70\times 10^{-14}$ & 33.98 & 5.59 & 12.43 & 2.55   & 7.18 \\
J1747$-$2958  &\ \ 2.52    & 0.0988 & $6.13\times 10^{-14}$& 36.40 & 4.41 & 12.40 & 4.38  & 8.08 \\
J1801$-$2451  &\ \ 3.80     & 0.1249 & $1.28\times 10^{-13}$& 36.41 & 4.19 & 12.61 & 4.23  & 8.14 \\
J1803$-$2137  &\ \ 4.40     & 0.1337 & $1.34\times 10^{-13}$& 36.34 & 4.20 & 12.63 & 4.23  & 8.12 \\
J1809$-$1917  &\ \ 3.27    & 0.0827 & $2.55\times 10^{-14}$& 36.26 & 4.71 & 12.17 & 4.39  & 7.97 \\
J1809$-$2332  &\ \ 0.88    & 0.1468  & $3.44\times 10^{-14}$ & 35.63 & 4.83 & 12.36 & 3.83  & 7.79 \\
\hline
\end{tabular}
\end{table*}
\begin{table*}
\renewcommand{\arraystretch}{1.38}
\contcaption{}
\begin{tabular}{lllllllll}
\\
\hline
\hline
Source name & Distance  &  $P$  & $\dot{P}$ &  $\log\dot{E}$ &  $\log \tau$  
&  $\log B_{\rm s}$  & $\log B_{\rm lc}$  &$\log\epc$ \\
& (kpc) & (s) &(s/s) & (erg/s) & (yr) & (G)& (G) & (statV/cm) \\
\hline
J1811$-$1925  &\ \ 5.00     & 0.0646 & 4.40$\times 10^{-14}$& 36.81 & 4.37 & 12.23 & 4.77  & 8.20 \\
J1813$-$1246  &\ \ 2.64    &0.0481  & 1.76$\times 10^{-14}$ & 36.79 & 4.64 & 11.97 & 4.89  & 8.13 \\
J1813$-$1749  &\ \ 4.70    & 0.0447 & 1.27$\times 10^{-13}$& 37.75 & 3.75 & 12.38 & 5.40  & 8.59 \\
J1826$-$1256  &\ \ 1.55    & 0.1102 & 1.21$\times 10^{-13}$ & 36.49 & 4.16 & 12.57 & 4.41 & 8.19 \\
J1826$-$1334  &\ \ 3.61    & 0.1010 & 7.53$\times 10^{-14}$& 36.45 & 4.33 & 12.45 & 4.40  & 8.12 \\
J1833$-$1034  &\ \ 4.10     & 0.0619 & 2.02$\times 10^{-13}$& 37.53 & 3.69 & 12.55 & 5.15  & 8.55 \\
J1836+5925  &\ \ 0.30    & 0.1733 & 1.50$\times 10^{-15}$ &34.04  & 6.26 & 11.71 & 2.97  & 7.04 \\
J1838$-$0537  &\ \ 1.30    & 0.1457 & 4.72$\times 10^{-13}$ &36.78  & 3.69 & 12.92 & 4.40  & 8.36 \\
J1838$-$0655  &\ \ 6.60    & 0.0705 & 4.93$\times 10^{-14}$& 36.74 & 4.36 & 12.28 & 4.70  & 8.18 \\
J1856+0113  &\ \ 3.30    & 0.2674 & 2.08$\times 10^{-13}$& 35.63 & 4.31 & 12.88 & 3.57  & 7.92 \\
J1907+0602  &\ \ 2.37    & 0.1066 & 8.68$\times 10^{-14}$ & 36.45 & 4.29 & 12.49 & 4.38  & 8.13 \\
J1930+1852  &\ \ 7.00   & 0.1369 & 7.51$\times 10^{-13}$& 37.08 & 3.46 & 13.01 & 4.57 & 8.49 \\
J1932+1059  &\ \ 0.31    & 0.2265 & 1.16$\times 10^{-15}$& 33.59 & 6.49 & 11.71 & 2.62  & 6.86 \\
J1952+3252  &\ \ 3.00    & 0.0395 & 5.84$\times 10^{-15}$& 36.57 & 5.03 & 11.69 & 4.87 & 7.97 \\
J1957+5033  &\ \ 1.37    & 0.3748  & 7.08$\times 10^{-15}$ & 33.72 & 5.92 & 12.22 & 2.47  & 7.04 \\
J2021+3651  &\ \ 1.80   & 0.1037 & 9.57$\times 10^{-14}$  & 36.53 & 4.24 & 12.50 & 4.43  & 8.16 \\
J2021+4026  &\ \ 2.15 & 0.2653  &5.47$\times 10^{-13}$ & 35.80 & 4.89 & 12.59 & 3.29  & 7.63 \\
J2022+3842  &\ \ 10.00   & 0.0486  & 8.61$\times 10^{-14}$ & 37.48 & 3.95 & 12.32 & 5.23 & 8.47 \\
J2043+2740  &\ \ 1.48   & 0.0961   &1.27$\times 10^{-15}$ & 34.75 & 6.08 & 11.55 & 3.57  & 7.25\\
J2055+2539  &\ \ 0.62   & 0.3196 & 4.08$\times 10^{-15}$ & 33.69 & 6.09 & 12.06 & 2.52  & 6.99 \\
J2225+6535  &\ \ 0.90    & 0.6825 & 9.66$\times 10^{-15}$& 33.08 & 6.05 & 12.41 & 1.88  & 6.84 \\
J2229+6114  &\ \ 3.00    & 0.0516 & 7.83$\times 10^{-14}$& 37.34 & 4.02 & 12.31 & 5.14  & 8.42 \\
J2337+6151  &\ \ 0.70  & 0.4954  & 1.93$\times 10^{-13}$ & 34.80 & 4.61 & 13.00 & 2.88  & 7.63 \\
\hline
\\
\end{tabular}
\end{table*}

\section{Correlations between non-thermal spectral properties and timing variables}

We first examine correlations between non-thermal X-ray spectral properties
and pulsar timing variables to check the consistency of this work and earlier literature
and to expand such kind of efforts for possible new results.
The non-thermal X-ray properties are characterized  by the luminosity, $\Lp$, 
in the energy range from 0.5 keV to 8 keV obtained from the power-law component in the best-fit spectral model,
and the photon power index, $\gmp$, of that power law. 
The timing variables considered in the literature to date include period $P$, period time-derivative $\pdot$, 
spin-down power $\edot$, characteristic age $\tau$,
the dipole magnetic field strength at the stellar surface $\bs$ and at the light cylinder $\blc$.  
Prompted by the discovery of $\blc$'s significant role in these correlations besides $\edot$ \citep{hsiang21}, 
in this work we introduce one more variable: $\epc=\dVh$, where
 $\dV$ is the potential difference between the magnetic pole and the rim of the polar cap for a pulsar in vacuum \citep{goldreich69}
and $h$ is the radius of the polar cap.
$\dV$ is about the highest potential drop that can be developed in a pulsar's magnetosphere
and therefore $\epc$ is of the order of the highest electric field strength that can occur in the polar gap.
$\epc$ is apparently related to acceleration of particles and can be very relevant.

These timing variables are all functions of $P$ and $\pdot$.
We adopt conventional values of the moment of inertia $I = 10^{45}$ g cm$^2$ and radius $R = 10$ km for
the pulsar, and have (all in gaussian units)
$\edot=I\Omega\dot{\Omega}=3.9\times 10^{46}P^{-3}\pdot$, 
$\tau=0.5P\pdot^{-1}$,
$\bs=3.2\times 10^{19}(P\pdot)^{0.5}$,
$\blc=\bs(R/R_{\rm lc})^3=2.9\times 10 ^8 P^{-2.5}\pdot^{0.5}$,
and 
$\epc=\dVh=4.9\times 10^{13} P^{-1}\pdot^{0.5}$,
where $\Omega=2\pi/P$, the light cylinder radius $R_{\rm lc}=c/\Omega$,
$\dV=\frac{1}{2}\left(\frac{R\Omega}{c}\right)^2R\bs$ ($\dV\propto\edot^{0.5}$) and
the polar cap radius $h=R\sqrt{\frac{R\Omega}{c}}$.

\begin{figure*}
\includegraphics[width=16cm]{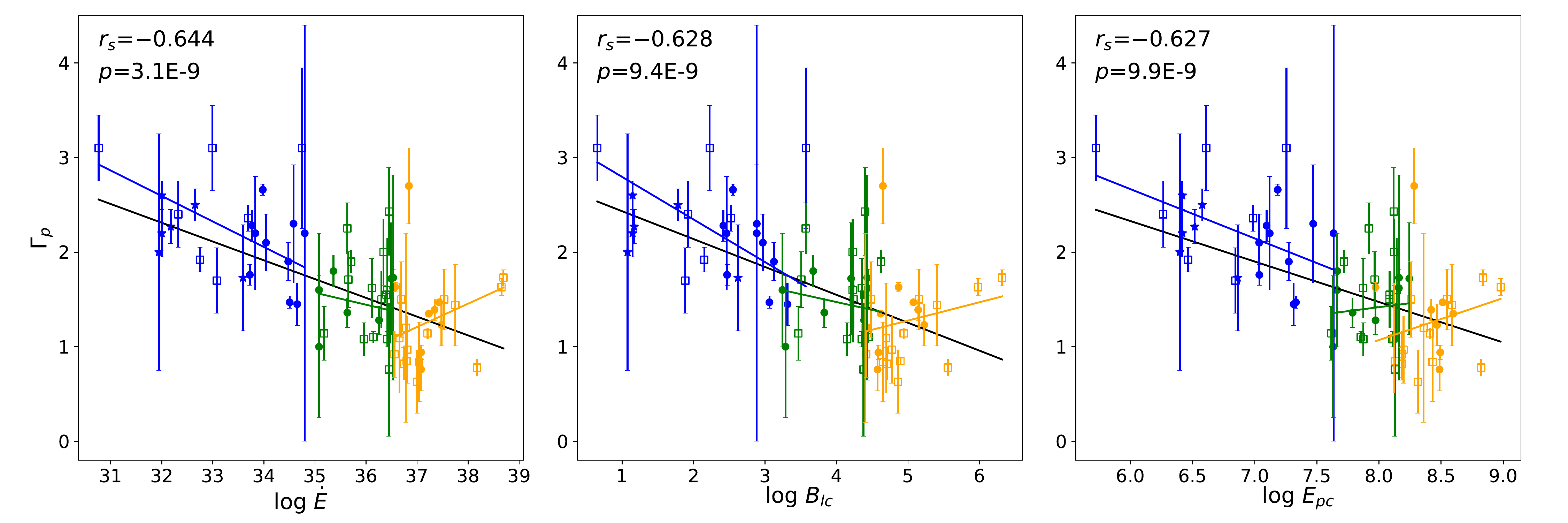}
\caption{The power index $\gmp$ of the non-thermal X-ray power-law component versus $\edot$, $\blc$ and $\epc$.
These are the three with the strongest correlations between $\gmp$ and timing variables.
See the caption in Figure \ref{fig:lptv} for other details.
}
\label{fig:gmtv}
\end{figure*}

\begin{figure}
\includegraphics[width=8cm]{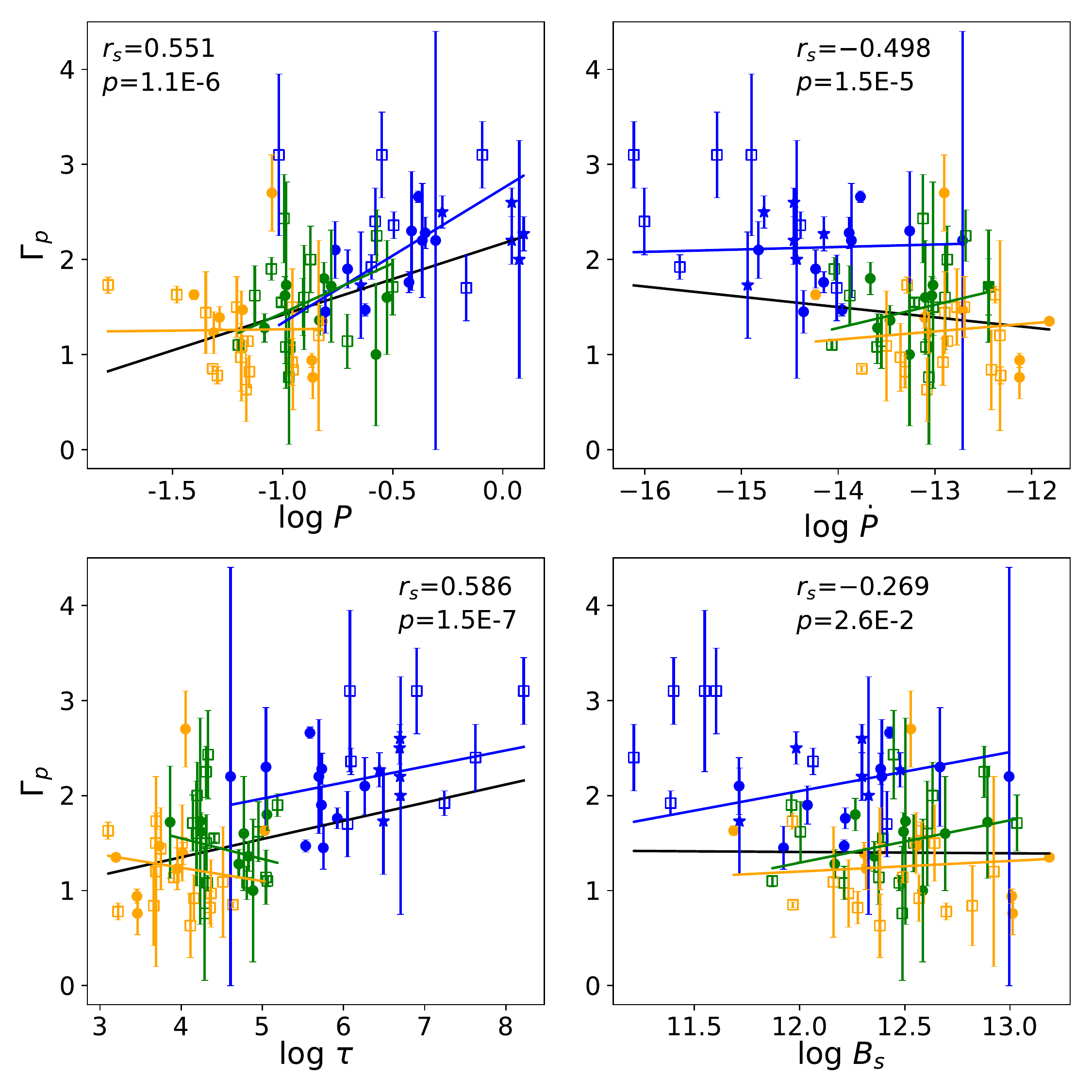}
\caption{The power index $\gmp$ of the non-thermal X-ray power-law component versus $P$, $\pdot$, $\tau$ and $B_{\rm s}$.
These are the four with weaker correlations between $\gmp$ and timing variables.
See the caption in Figure \ref{fig:lptvx} for other details.
}
\label{fig:gmtvx}
\end{figure}

$\Lp$ and $\gmp$ versus all the timing variables are plotted in Figures \ref{fig:lptv} -- \ref{fig:gmtvx}.
The Spearman rank-order correlation coefficient $r_s$ is employed  to indicate the significance of possible correlations.
Each $r_s$ and its associated random probability $p$ are noted in the figures. 
As found earlier, $\Lp$ is very strongly correlated with $\edot$ as well as $\blc$ \citep{hsiang21}.
We found, in addition, $\Lp$'s correlation with $\epc$ is equally strong.
This again suggests that $\edot$ should not be considered as the only major factor determining the non-thermal X-ray emissions from pulsars.
We note that $\Lp$'s correlations with the pulsar spin period $P$ and the characteristic age $\tau$ are also considerably strong.


The best fits to the three strongest correlations, using the least-square method to fit a linear function of the logarithms of these variables,  are
\beq
\Lp\propto\edot^{0.88\pm 0.06}\,\,\,\,\, (\chi^2_\nu=3.98)\,\,\, ,
\label{eq:lpedot}
\eeq
\beq
\Lp\propto\blc^{1.24\pm 0.09}\,\,\,\,\, (\chi^2_\nu=4.15)\,\,\, ,
\label{eq:lpblc}
\eeq
and
\beq
\Lp\propto\epc^{2.16\pm 0.16}\,\,\,\,\, (\chi^2_\nu=4.36)\,\,\, .
\label{eq:lpepc}
\eeq
The fitting uncertainties reported here and hereafter are all at a confidence level of 68\%.
Similar to what were found in earlier literature (e.g. \citet{possenti02,li08,hsiang21}),
the reduced chi-squares, $\chi^2_\nu$, of these best fits are large and statistically not acceptable.
We note that, in this work, best fits and $\chi^2_\nu$ are only suggestive, because the quoted uncertainties
contain distance uncertainties and are not derived rigorously with proper statistics. 
Since timing variables are all functions of $P$ and $\pdot$, 
a two-variable fitting was conducted, that is, to fit
the luminosity logarithm as a linear function of $\log P$ and $\log\pdot$ directly.
The best fit is
\beq
\Lp\propto P^{-2.84\pm 0.30}\pdot^{0.79\pm 0.14}\,\,\,\,\, (\chi^2_\nu=4.00) \,\,\,\, .
\label{eq:lpppdot}
\eeq
The fitting goodness here is similar to that reported earlier \citep{possenti02,hsiang21}.
One may turn this best fit into functions of any two timing variables, that is,
$\Lp\propto\edot^{0.56}\blc^{0.47}$, 
$\Lp\propto\edot^{1.26}\epc^{-0.94}$, or
$\Lp\propto\blc^{0.84}\epc^{0.74}$.

In the past, $\gmp$ was not found to correlate with any timing variables.
Now with a larger sample, we found that it is in fact correlated with all the timing variables except for $\bs$
in a way similar to that of $\Lp$ but with a somewhat weaker significance (Figure \ref{fig:gmtv}).
The best fits to the three strongest correlations, using the least-square method to fit $\gmp$ as a linear function of the logarithms of these variables,  are
\beq
\gmp=\log\edot^{-0.20\pm 0.03}+(8.66\pm 1.26)\,\,\,\,\, (\chi^2_\nu=13.9)\,\,\, ,
\label{eq:gmedot}
\eeq
\beq
\gmp=\log\blc^{-0.29\pm 0.05}+(2.73\pm 0.22)\,\,\,\,\, (\chi^2_\nu=13.3)\,\,\, ,
\label{eq:gmblc}
\eeq
and
\beq
\gmp=\log\epc^{-0.43\pm 0.09}+(4.90\pm 0.70)\,\,\,\,\, (\chi^2_\nu=15.3)\,\,\, .
\label{eq:gmepc}
\eeq
The $\chi^2_\nu$ of these best fits are all much larger than that of $\Lp$, 
indicating a larger scatter in the correlations.
The dependence of $\gmp$ on timing variables is also weaker.
The best fit for the fitting directly with $P$ and $\pdot$ is
\beq
\gmp=\log (P^{0.77\pm 0.13}\pdot^{-0.13\pm 0.05})+ (0.54\pm 0.64)
\label{eq:gmppdot}
\eeq
with $\chi^2_\nu=13.3$.
This can also be turned into functions of any two of $\edot$, $\blc$ and $\epc$.
Although in Eqs(\ref{eq:gmedot})-(\ref{eq:gmepc}) the best fit $\chi^2_\nu$ is large and
the dependence of $\gmp$ on $\edot$, $\blc$ and $\epc$ seems weak,
their correlations are clearly shown in Figure \ref{fig:gmtv}, with
a random probability of order of $10^{-9}$, based on the Spearman's rank-order correlation coefficient.
This is reported in the literature for the first time.

With a large sample of 68 pulsars, we divided them into three groups according to their spin-down power.
Members of different groups are color-coded in Figures \ref{fig:lptv} -- \ref{fig:gmtvx}.
The best fits for each groups are also plotted.
We note that, in the correlations of $\Lp$ versus $\edot$, $\blc$ and $\epc$, slopes of the best fits are larger
for higher-power groups.  
There is also a tendency of increasing slopes with higher spin-down-power groups for $\gmp$,
 from negative to positive values. 
These slopes (i.e. power indices of $\edot$, $\blc$ and $\epc$) 
as shown in Figure \ref{fig:lptv} and Figure \ref{fig:gmtv} are listed in Table \ref{tab:grp} for readers' reference.
The power indices  in Eqs.(\ref{eq:lpedot})--(\ref{eq:lpepc}) and in Eqs.(\ref{eq:gmedot})--(\ref{eq:gmepc}) are 
the combined results of these three groups.
The stronger dependence of $\Lp$ on these three timing variables for higher spin-down-power groups
is obvious. 
\begin{center}
\begin{table*}
\caption{The power index $\alpha$ in the relationship $\Lp\propto V^\alpha$ for the three spin-down-power groups, 
where $V$ stands for $\edot$, $\blc$ or $\epc$. The numbers in the parentheses are $\chi^2_\nu$ of the corresponding fits.}
\label{tab:grp}
\begin{tabular}{llll}
\hline
\hline
V & Low power group & Median power group& High power group \\
\hline
  & & $\Lp\propto V^\alpha$ & \\
 \hline
$\edot$  &  $0.35\pm 0.09$ (0.91)  &  $1.09\pm 0.35$ (2.80)  & $1.70\pm 0.32$ (4.41)     \\
$\blc$  &   $0.45\pm 0.14$ (1.03)  &  $1.27\pm 0.41$ (2.81)  & $1.94\pm 0.44$ (5.50)     \\
$\epc$  &   $0.86\pm 0.22$ (0.88)  &  $2.00\pm 0.94$ (3.41)  & $3.89\pm 0.84$ (5.21)     \\
\hline
 & & $\gmp=\log V^\alpha$ plus a constant &  \\
 \hline
$\edot$  &  $-0.27\pm 0.12$ (9.23)  &  $-0.13\pm 0.21$ (5.15)  & $0.25\pm 0.13$ (17.5)     \\
$\blc$  &   $-0.45\pm 0.16$ (8.29)  &  $-0.17\pm 0.28$ (5.15)  & $0.19\pm 0.18$ (19.5)     \\
$\epc$  &   $-0.52\pm 0.31$ (10.0)  &  $0.17\pm 0.44$ (5.21)  & $0.44\pm 0.24$ (17.7)     \\
\hline
\end{tabular}
\end{table*}
\end{center}

\section{Correlations between non-thermal and thermal emissions}
As mentioned in the introduction section, there are good reasons to expect 
some connection between non-thermal and thermal X-ray emissions from pulsars.
In the sample of 68 pulsars that we study in this work, thermal emissions are detected in 32 pulsars
together with their non-thermal emissions, described by a PL+BB spectral model.
These 32 pulsars do not show any distinct behavior in their timing variable distribution from the others that do not show thermal emissions.
This is shown  in Figures \ref{fig:lptv} -- \ref{fig:gmtvx}, in which these 32 are plotted with solid symbols and the other 36 with open ones.

\begin{figure}
\includegraphics[width=8cm]{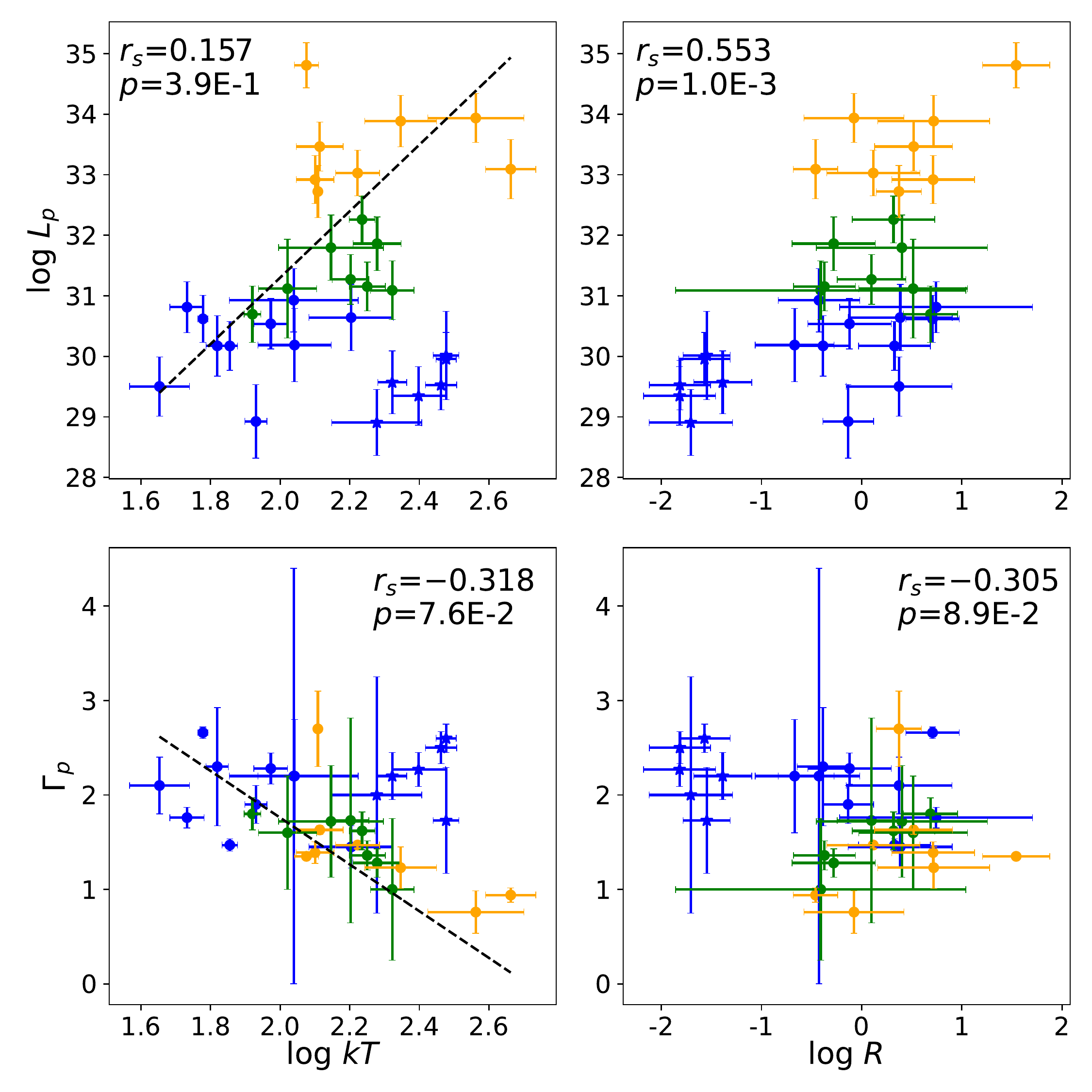}
\caption{
$\Lp$ and $\gmp$ versus  $kT$ and $R$, respectively.
The six pulsars with a hot spot are plotted with a star symbol.
In the legend, $r_s$ is the Spearman rank-order correlation coefficient and $p$ is the corresponding random probability for the whole sample of
32 pulsars, i.e., including the six with a hot spot.
The two straight, dashed lines in the left panels are the best fits described in Eq.(\ref{eq:lpkt26}) and Eq.(\ref{eq:gmkt26}) for the 26 pulsars, i.e., excluding the six with a hot spot.
The color code here and in all the following figures means the same thing as that in previous figures.
}
\label{fig:lpgmtr}
\end{figure}

The correlations between the non-thermal properties, $\Lp$ and $\gmp$, and the thermal ones, 
temperature $kT$ (in units of eV) and the emitting-region radius $R$ (in units of km), 
are shown in Figure \ref{fig:lpgmtr}.
One can see that these correlations are weak.
The only stronger correlations appear between $\Lp$ and $kT$ and between $\gmp$ and $kT$ when
the six pulsars (in fact five: PSR J0946+0951 is counted twice for its different modes) with the so-called hot spots are excluded.
Their Spearman's rank-order correlation coefficients are 0.71  and $-0.75$, corresponding to a $p$-value of 
 $4.6\times 10^{-5}$ and $1.1\times 10^{-5}$, respectively. 
The data point scatter is, however, still very large.
The best fit of $\Lp$ and $\gmp$ as a function of $kT$ for the sample of 26 pulsars are
\beq
\Lp\propto (kT)^{5.48\pm 1.24}\,\,\,\,\, (\chi^2_\nu=5.65)
\label{eq:lpkt26}
\eeq
and
\beq
\gmp=\log(kT)^{-2.47\pm 0.47}+(6.71\pm 0.94)\,\,\,\,\, (\chi^2_\nu=6.41)\,\,\, .
\label{eq:gmkt26}
\eeq

Since the $\chi^2_\nu$ values in the above are large, we started to look for possible, tighter relationships 
in a multi-dimensional space, similar to what we did to obtain Eq.(\ref{eq:lpppdot}) and Eq.(\ref{eq:gmppdot}).
This is also equivalent to finding a fundamental plane in a multi-dimensional space.
Adding $P$ and $\pdot$ to be independent variables together with $kT$, still with the sample of 26 pulsars,
the best fit for $\Lp$ and $\gmp$ yields a $\chi^2_\nu$ value of 2.85 and 6.26, respectively.
This fitting improvement apparently is not significant.
The relationship reported earlier in \citet{hsiang21} between $\gmp$ and $kT$, $P$ and $\pdot$,
obtained from a sample of 12 pulsars, is not confirmed by the current, larger sample.

\begin{figure}
\includegraphics[width=8cm]{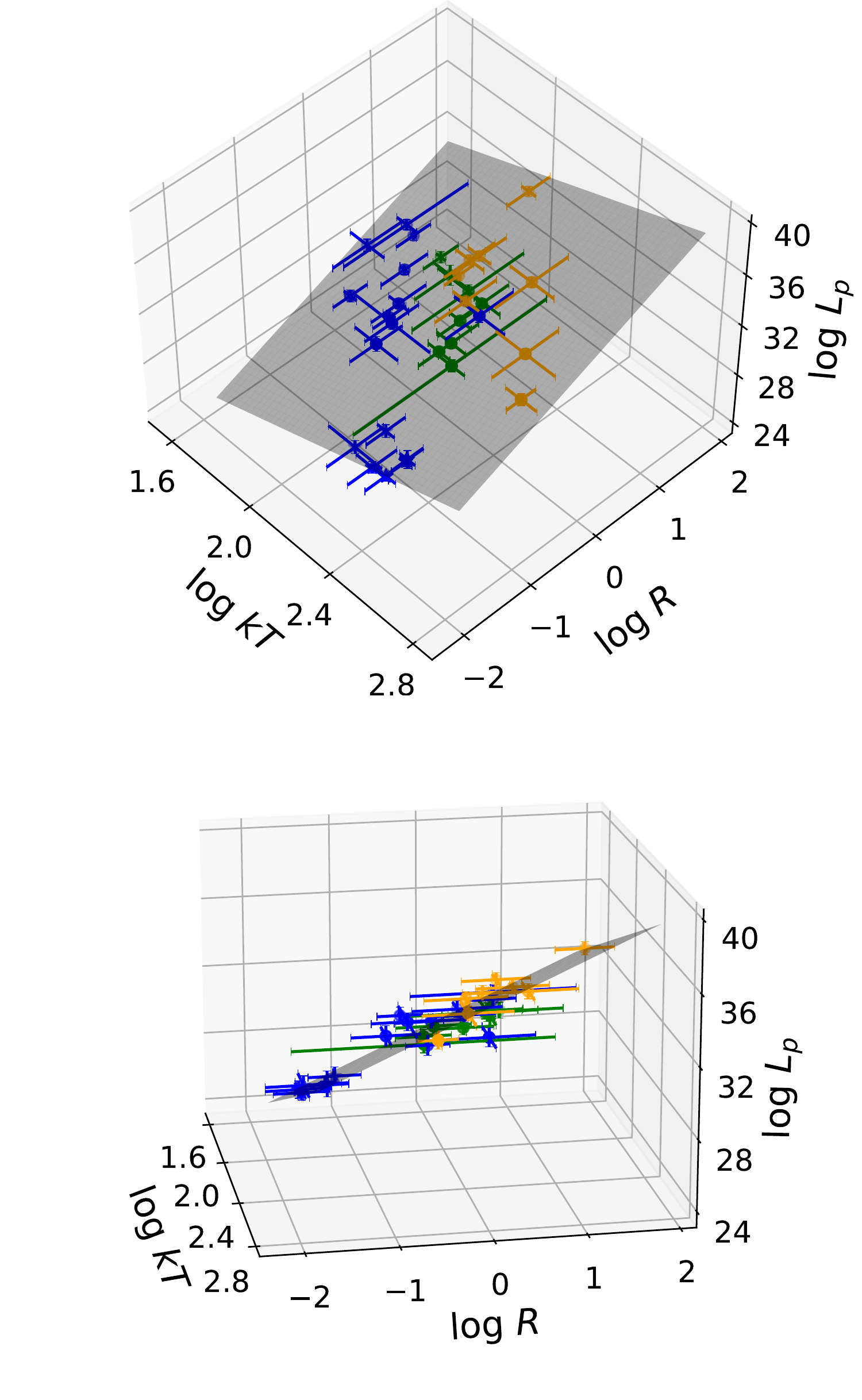}
\caption{
The fundamental plane in the space of \{$\Lp$, $kT$, $R$\}.
The upper panel is a more face-on view to show the separate group of the six hot-spot pulsars.
The lower one is a more edge-on view to show the plane.
}
\label{fig:lptr3d}
\end{figure}

\begin{figure}
\includegraphics[width=8cm]{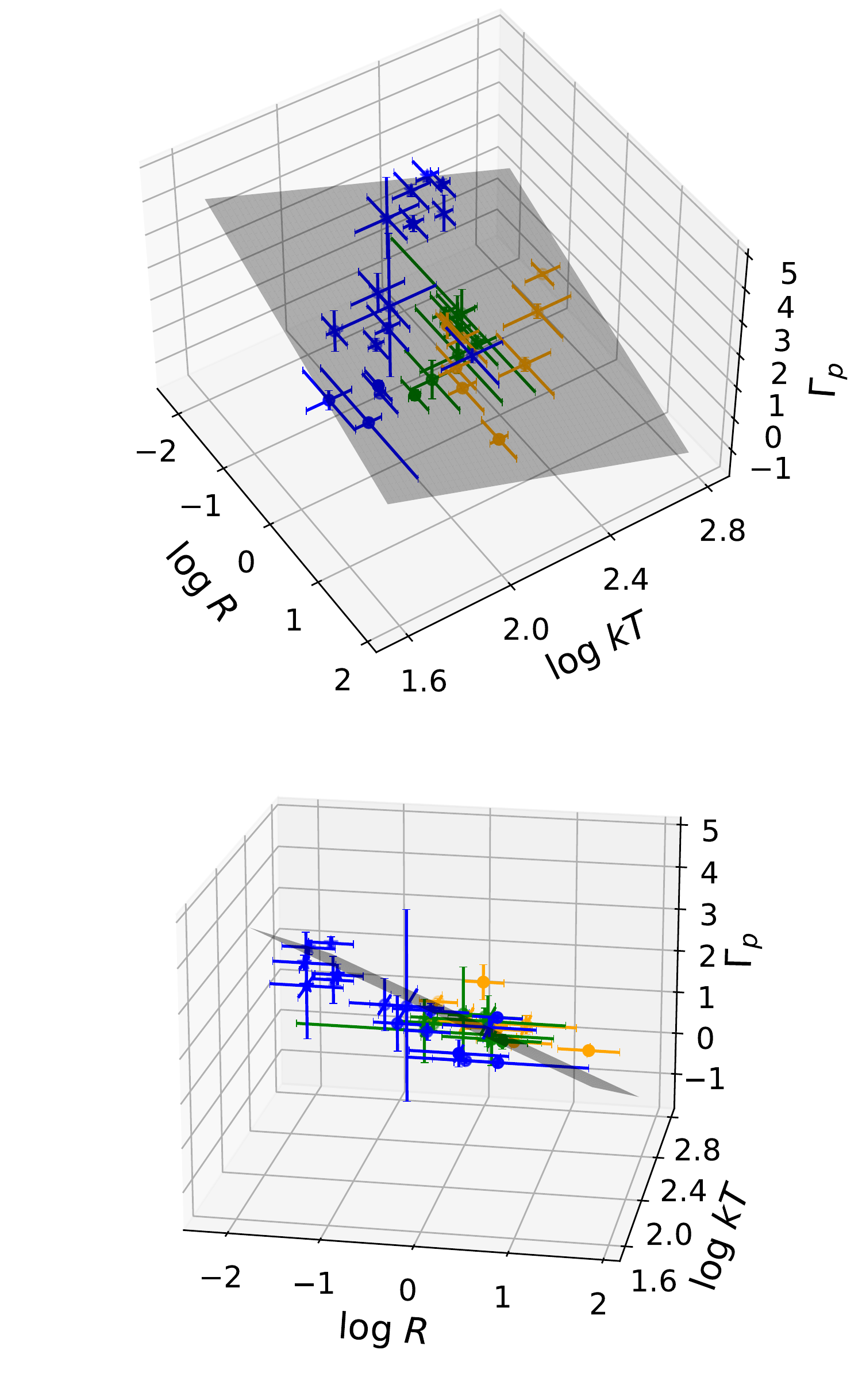}
\caption{
The fundamental plane in the space  of \{$\gmp$, $kT$, $R$\}.
The upper panel is a more face-on view to show the separate group of the six hot-spot pulsars.
The lower one is a more edge-on view to show the plane.
}
\label{fig:gmtr3d}
\end{figure}

We then considered the space spanned by $kT$, $R$ and one of $\Lp$ and $\gmp$.
Much tighter relationships, indicated by a small $\chi^2_\nu$, were found.
They are
\beq
\Lp\propto (kT)^{5.95\pm 0.77}R^{2.57\pm 0.37}\,\,\,\,\, (\chi^2_\nu=0.58)
\label{eq:lpktr26}
\eeq
and
\beqary
\gmp &=& \log((kT)^{-2.65\pm 0.57}R^{-1.04\pm 0.37})\nonumber \\
& & +(7.56\pm 1.27)\,\,\,\,\,\, (\chi^2_\nu=1.46) .
\label{eq:gmktr26}
\eeqary
In fact, we found that, in the space of \{$\Lp$, $kT$, $R$\} and of \{$\gmp$, $kT$, $R$\}, the sample of 32 pulsars (i.e. including those six with hot spots) has similar
relations:
\beq
\Lp\propto (kT)^{5.96\pm 0.64}R^{2.24\pm 0.18}\,\,\,\,\, (\chi^2_\nu=0.49)
\label{eq:lpktr32}
\eeq
and
\beqary
\gmp &= &\log((kT)^{-2.56\pm 0.44}R^{-0.85\pm 0.14})\nonumber \\
& & +(7.29\pm 0.95) \,\,\,\,\,\, (\chi^2_\nu=1.34) .
\label{eq:gmktr32}
\eeqary
This may be an indication that properties of non-thermal emissions are linked to thermal emissions, 
no matter whether there is a hot spot or not.
These fundamental planes are plotted in Figures \ref{fig:lptr3d} and \ref{fig:gmtr3d}, in which one can see that the six hot-spot pulsars
form a separate group but still are located in the same fundamental plane with the other 26.
 
Although these relations are tight (not quite so for Eq.(\ref{eq:gmktr26}) and Eq.(\ref{eq:gmktr32})),
timing variables might still play some role in determining properties of non-thermal X-ray emissions.
We therefore included $P$ and $\pdot$ in the fitting and found the best fit (for the sample of 32 pulsars) to be
\beq
\Lp\propto (kT)^{6.17\pm 1.14}R^{2.38\pm 0.48}P^{0.48\pm 0.59}\pdot^{0.09\pm 0.27}\,\, (\chi^2_\nu=0.50)
\label{eq:lpktrpp32}
\eeq
and
\beqary
\gmp &= &\log((kT)^{-5.80\pm 1.93}R^{-2.29\pm 0.85}P^{-1.19\pm 0.88}\pdot^{0.94\pm 0.44})\nonumber \\
 & & +(26.0\pm 9.23)\,\,\,\,(\chi^2_\nu=0.84) .
\label{eq:gmktrpp32}
\eeqary
The dependence of $\Lp$ on $kT$ and $R$ does not change much from the space of 
\{$\Lp$, $kT$, $R$\} to the space of \{$\Lp$, $kT$, $R$, $P$, $\pdot$\}.
As shown in Eq.(\ref{eq:lpktrpp32}) the dependence on $P$ and $\pdot$ is quite weak.
On the other hand, the dependence of $\gmp$ on $kT$ and $R$ changes a lot.
Timing variables play a more significant role than for the case of $\Lp$.
This fitting has a small $\chi^2_\nu$ value of 0.84.
As noted earlier, we are not able to demonstrate in a more statistically rigorous way
whether the $\chi^2_\nu$ values in Eq.(\ref{eq:lpktrpp32}) and Eq.(\ref{eq:gmktrpp32}) indicate
a statistically acceptable fit or not,
because the uncertainty to pulsar distances is assigned somewhat arbitrarily at the level of 40\%.
Nonetheless, the best fits with a smaller $\chi^2_\nu$ do provide intriguing guidances for theoretical modeling.  

\section{Conclusions and Discussion}

We employed a large sample of 68 pulsars to study the connections of their non-thermal X-ray properties,
the power-law component luminosity $\Lp$ in 0.5-8 keV and the spectral photon power index $\gmp$,
with their thermal emission properties, the surface temperature $kT$ and the emitting-region radius $R$,
and  their various timing variables.
We found the following:
(i) $\Lp$ is strongly correlated with $\edot$, $\blc$ and $\epc$ at a similar significance level (Figure \ref{fig:lptv}). 
The strong correlation with $\blc$ has been reported in \citet{malov19} and \citet{hsiang21},
but the one with $\epc$ is reported here for the first time. 
It suggests that the spin-down power $\edot$ may not be the only major factor determining $\Lp$.
(ii) The dependence of  $\Lp$ on $\edot$, $\blc$ and $\epc$ shows a clear trend of being
stronger with increasing $\edot$ (Figure \ref{fig:lptv}).
(iii)  $\gmp$ is also strongly correlated with $\edot$, $\blc$ and $\epc$ at a similar significance level, 
but somewhat weaker than that of $\Lp$ (Figure \ref{fig:gmtv}). This is the first time that this correlation is reported in the literature.
(iv) Similar to $\Lp$, the dependence of   $\gmp$ on $\edot$, $\blc$ and $\epc$ changes with the spin-down power.
It turns from a negative correlation for the lower power group to a positive one for the high power group (Figure \ref{fig:gmtv}).
(v) Relatively tight correlations are found among $\Lp$, $kT$ and $R$ and among $\gmp$, $kT$ and $R$ 
(Eqs.(\ref{eq:lpktr32})--(\ref{eq:gmktr32}) and Figures \ref{fig:lptr3d} and \ref{fig:gmtr3d}).
These are strong indication of the connection between thermal and non-thermal X-ray emissions.
A tighter fundamental-plane relation for $\gmp$ is found when two more dimensions, $P$ and $\pdot$, are
included.  

Items (i)--(iv) above provide finer constraints to theoretical modeling.
In the past, although hampered by the sample size and by mixtures with emissions from pulsar wind nebulae and with thermal emissions,
a strong correlation between non-thermal X-ray emission from pulsars and their spin-down power was found and
the dependence is roughly linear with $\alpha$ in $\Lp\propto\edot^\alpha$ being in the range from 0.9 to 1.4 
(see the introduction in \citet{hsiang21} for a brief review).
An earlier outer-gap model predicts $\Lp\propto\edot^{1.15}$ \citep{cheng04}.
From the results presented in this paper, however, one sees that
$\blc$ and $\epc$ may play an equally important role as $\edot$ in determining
the properties of non-thermal X-ray emissions from pulsars.
The changing dependence with different spin-down-power groups
also indicates that somewhat different mechanisms may be at work for different energetics regimes.

Non-thermal X-rays are believed to originate from pair plasma created in pulsars' magnetosphere.
Pair production are expected to occur either near the polar cap or in the so-called outer-gap regions
in different models. They are, however, both related to thermal emissions from the stellar surface.
Near the polar cap, charged particles are accelerated in the polar gap to emit high-energy photons, which make
pairs in the strong magnetic field. The acceleration of those charged particles does not only depend on the
available potential drop in the polar gap but also on the thermal X-ray photon bath, which may decelerate those 
charged particles through inverse Compton scattering. In the outer gap region, surface thermal X-ray photons
collide with seed high-energy photons to make pairs.
It is therefore expected that thermal emission is related to non-thermal X-rays.
The large scatter showing up in Figure \ref{fig:lptv} and Figure \ref{fig:gmtv} and in all the earlier studies in the literature
may indicate that, beside the geometry factor, $P$ and $\pdot$ are not enough to describe the properties of non-thermal X-rays.
Properties of thermal emissions probably should come into play.
Indeed, as mentioned in item (v) in the above, 
tight relations describing fundamental planes in the space of  \{$\Lp$, $kT$, $R$\} and of \{$\gmp$, $kT$, $R$\} are found.
Finally, $\Lp$ and $\gmp$ as a function of $kT$, $R$, $P$ and $\pdot$ are presented in Eqs.(\ref{eq:lpktrpp32}) and (\ref{eq:gmktrpp32}).

\begin{figure}
\includegraphics[width=8cm]{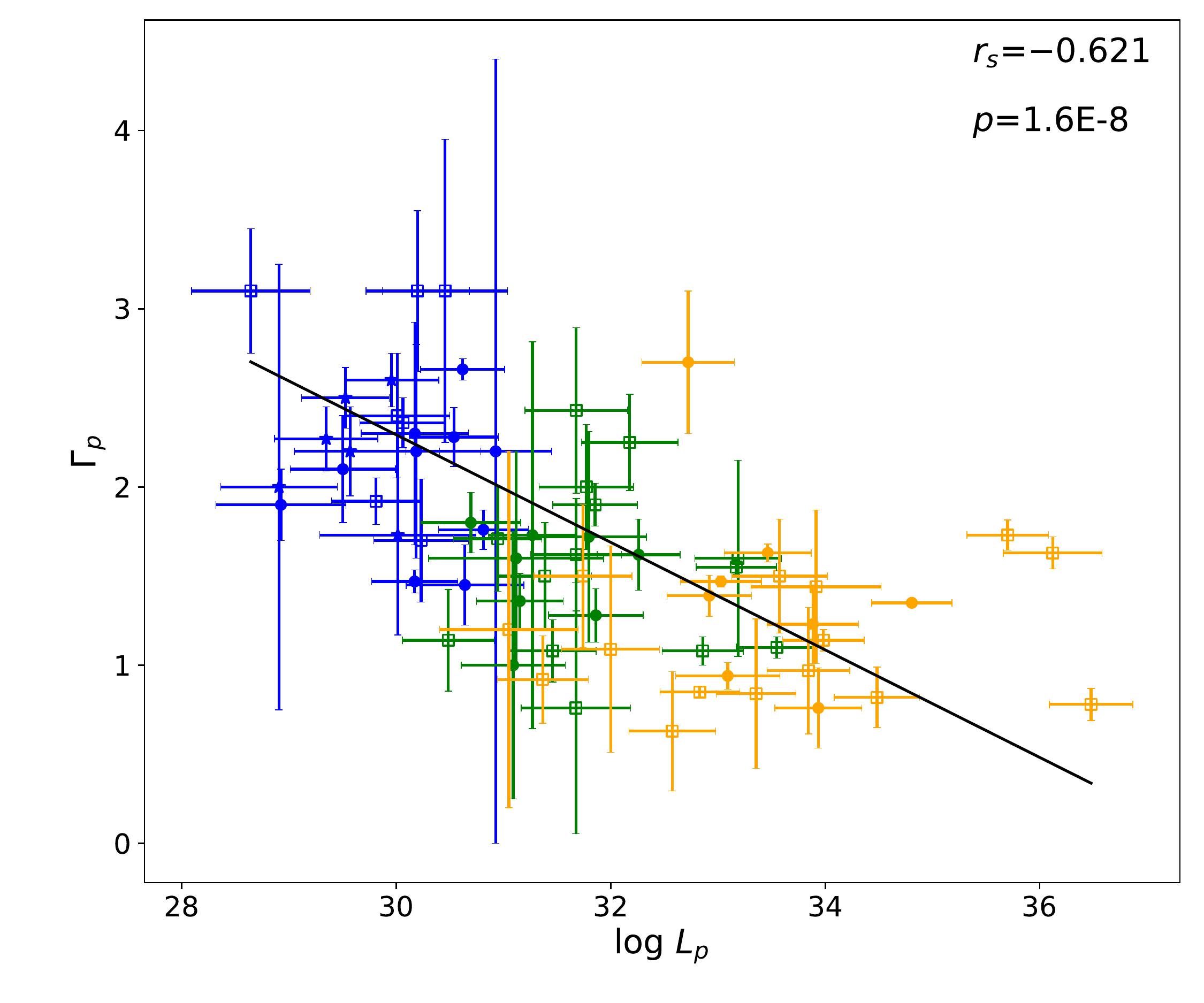}
\caption{
$\gmp$ versus $\Lp$.
In the legend, $r_s$ is the Spearman rank-order correlation coefficient and $p$ is the corresponding random probability for the whole sample of
68 pulsars.
The straight line is the best fit described in Eq.(\ref{eq:gmlp})
}
\label{fig:gmlp}
\end{figure}   
The power index $\gmp$ is also correlated with the non-thermal luminosity $\Lp$ 
at a significance level of the random probability being order of $10^{-8}$ (Figure \ref{fig:gmlp}).
The best fit to their relation is
\beq
\gmp=\log\Lp^{-0.30\pm 0.04}+(11.3\pm 1.2)\,\,\,\,\, (\chi^2_\nu=5.6)\,\,\, .
\label{eq:gmlp}
\eeq
This relation (the power index -0.30 of $\Lp$) for the sample of 68 pulsars
 is consistent with Eq.(\ref{eq:lpktr32}) and Eq.(\ref{eq:gmktr32}),
which are for the sample of 32 pulsars with thermal emissions.
In the search for a fundamental plane relation,
we did not try to do that in a space containing dimensions of both $\Lp$ and $\gmp$,
because these two quantities are two distinct dependent variables, which depend on independent variables 
$kT$ and $R$, and likely also $P$ and $\pdot$.    
Nonetheless, the trend of $\gmp$ decreasing with increasing $\Lp$, as described by Eq.(\ref{eq:gmlp}) and shown in Figure \ref{fig:gmlp}, 
is clear.
 
\begin{figure}
\includegraphics[width=8cm]{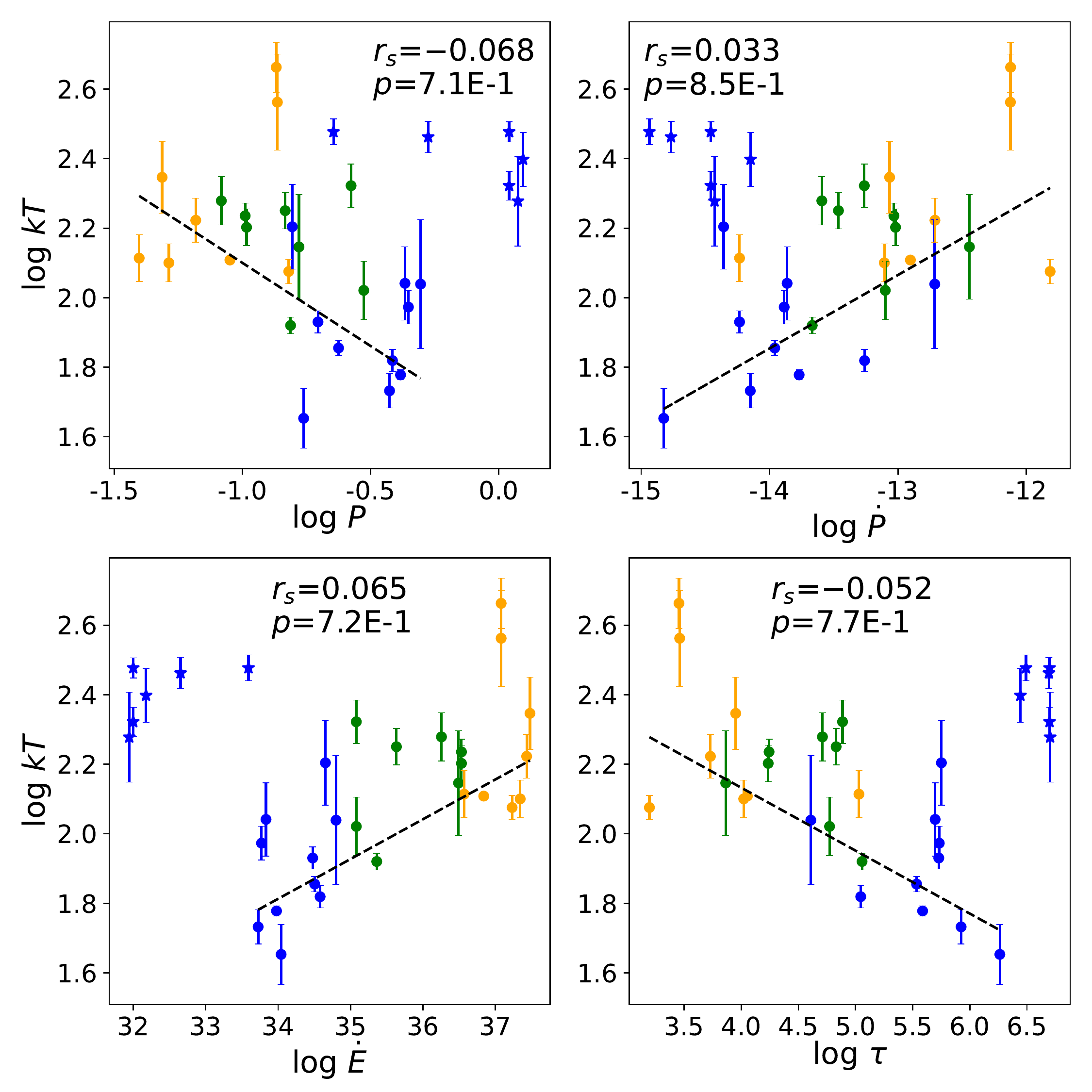}
\caption{The surface temperature $kT$ versus $P$, $\pdot$, $\edot$ and $\tau$.
$r_s$ and $p$ in the legend the Spearman rank-order correlation coefficient 
and  the corresponding random probability for the sample of
32 pulsars, that is, the six hot-spot pulsars plotted with a star symbol are included.
When those six are excluded, $r_s$ and $p$ are \{$-0.612$, $8.7\times 10^{-4}$\}, 
\{$0.522$, $6.2\times 10^{-3}$\}, \{$0.699$, $7.0\times 10^{-5}$\} and \{$-0.660$, $2.4\times 10^{-4}$\}
for the $P$, $\pdot$, $\edot$ and $\tau$ panels, respectively.
Dashed lines are the best fits to the 26 pulsars.
}
\label{fig:ttv}
\end{figure}   
\begin{figure}
\includegraphics[width=8cm]{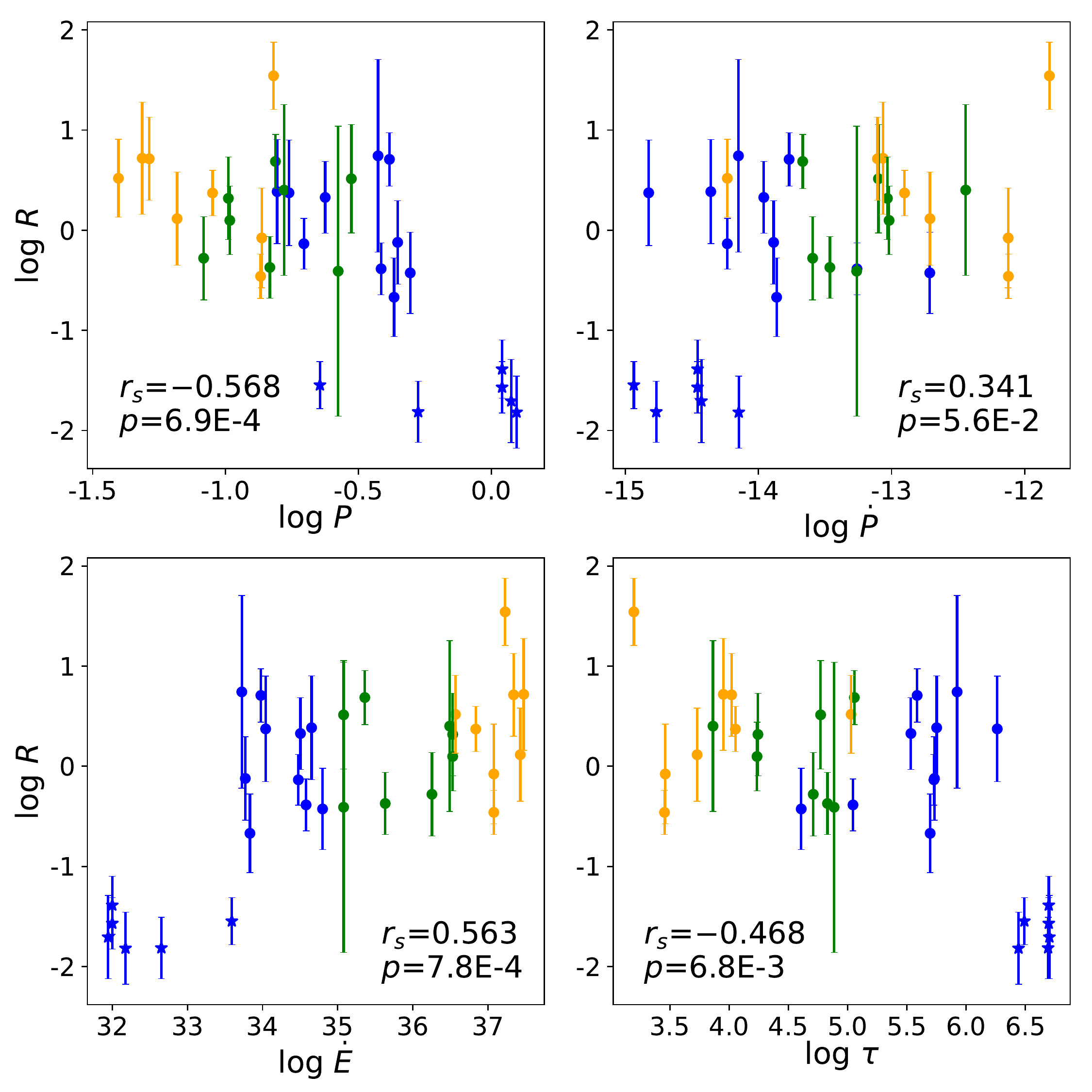}
\caption{The emitting-region radius $R$ versus $P$, $\pdot$, $\edot$ and $\tau$.
In the legend, $r_s$ is the Spearman rank-order correlation coefficient and $p$ is the corresponding random probability for the sample of
32 pulsars, that is, the six hot-spot pulsars plotted with a star symbol are included.
The six hot-spot pulsars clearly form a separate group.
}
\label{fig:rtv}
\end{figure}   

\begin{figure}
\includegraphics[width=8cm]{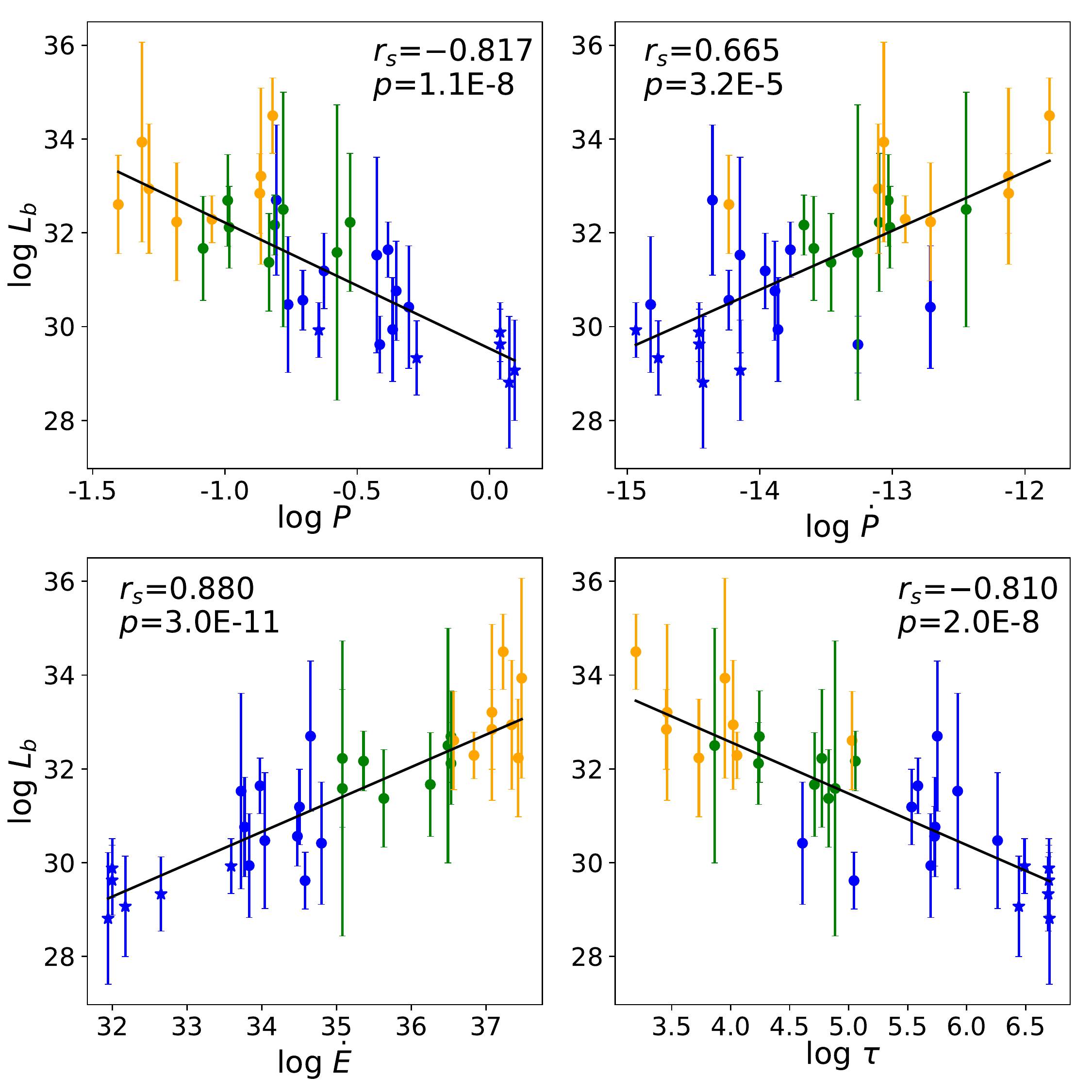}
\caption{Thermal luminosity $L_{\rm b}$ versus $P$, $\pdot$, $\edot$ and $\tau$.
The straight lines in black and $r_s$ and $p$ in the legend are the best linear fits, 
the Spearman rank-order correlation coefficient and  the corresponding random probability for the whole sample of
32 pulsars, respectively.
The slopes of those straight lines are $-2.68\pm 0.44$ for the upper left panel, $1.26\pm 0.21$ for upper right, $0.69\pm 0.08$ for lower left and
 $-1.10\pm 0.13$ for the lower right.
}
\label{fig:lbtv}
\end{figure}   

One may also wonder whether $kT$ and $R$ are just functions of $P$ and $\pdot$ and
the non-thermal X-ray properties can therefore be described by the two variables $P$ and $\pdot$
as in Eq.(\ref{eq:lpppdot}) and Eq.(\ref{eq:gmppdot}), for which the large $\chi^2_\nu$ values can be 
attributed solely to geometry factors. 
In Figures \ref{fig:ttv} and \ref{fig:rtv}, $kT$ and $R$ versus   $P$, $\pdot$, $\edot$ and $\tau$ are plotted.
Generally speaking, $kT$ and $R$ are not correlated with those timing variables,
except that, when the six hot-spot pulsars are excluded, $kT$ shows modestly strong correlations
with them. The scatter, however, is large.
On the other hand, if we consider the bolometric thermal luminosity $\Lb$, which is proportional to $R^2T^4$,
relatively strong correlations show up, as shown in Figure \ref{fig:lbtv}.
The six hot-spot pulsars follow the same correlations with the others, unlike that shown in Figures \ref{fig:ttv} and \ref{fig:rtv}.
It may suggest that $\Lb$ is a more relevant parameter to use than $R$.
We note that, in Eqs.(\ref{eq:lpktr32}) and (\ref{eq:gmktr32}), the dependence on $kT$ and $R$ cannot be replaced by $\Lb$ alone.
It is needed and equivalent to use any two among $kT$, $R$ and $\Lb$.

Although it is not within the scope of this paper, 
we note that $\Lb$ seems  to describe neutron star cooling better than the surface temperature
(see the lower right panels in Figures \ref{fig:ttv} and \ref{fig:lbtv}).
The temperature $T$ and the emitting-region radius $R$ obtained from spectral fitting with a blackbody component are
an averaged description of the surface thermal emission, which comes from a surface with a non-uniform temperature distribution.
Neutron cooling, through emission of neutrinos and photons, apparently does not only depend on the age $\tau$ 
but also on the neutron star mass, composition (at the core and the crust),
the state of matter, magnetic configuration (possible presence of higher multipoles) and so on (e.g. \citet{page09,potekhin15,potekhin20}).
The scatter in the relation of $\Lb$ versus $\tau$ is therefore inevitable.
More samples and more precise measurements will not eliminate that scatter, but may yield a more stable, ensemble-averaged  relation between $\Lb$ and $\tau$.
As far as the non-thermal emission is concerned, when $T$ and $\Lb$ (can be represented by $T$ and $R$) are determined,
those physical factors like mass, composition and the equation of state do not affect the production of
non-thermal X-ray emissions.
However, the possible presence of magnetic higher multipoles and the geometric factors (magnetic inclination and observer's viewing angles)
will still play significant roles in the search for fundamental planes in the space of \{$\Lp$, $kT$, $R$, $P$, $\pdot$\} and of \{$\gmp$, $kT$, $R$, $P$, $\pdot$\}.
We expect that, with more samples and more precise measurements, just like $\Lb$ versus $\tau$, 
we can only obtain a more stable, ensemble-averaged fundamental-plane relation.
The scatter around that relation is inevitable.
Nonetheless, such an ensemble-averaged relation is valuable for constraining theoretical models for non-thermal emissions from pulsars.


\section*{Acknowledgements}
This work is supported by the National Science and Technology Council (NSTC) of the Republic of China (Taiwan) under grants
MOST 110-2112-M-007-020 and MOST 111-2112-M-007-019.



\section*{Data availability}
The data employed by this article are available in the article and in the quoted references.


\label{lastpage}
\end{document}